\documentclass[aps,pre,10pt,showpacs,floatfix,onecolumn,nofootinbib,a4paper]{revtex4-2}

\usepackage{amssymb, amsmath, amsthm, mathtools}
\usepackage{bm}
\usepackage{mathrsfs}
\usepackage{hyperref,bookmark}
\usepackage{graphicx}
\usepackage{epsfig,color}
\usepackage{float}
\usepackage{subcaption}
\usepackage{verbatim}
\usepackage{dcolumn}
\usepackage{tikz}
\usepackage{url}

\usepackage[utf8x]{inputenc}
\usepackage[english]{babel}
\usepackage{mathrsfs}
\usepackage{epsfig}
\usepackage{listings}

\usepackage{mdwlist}
\usepackage{bm}
\usepackage{dsfont}
\usepackage{oldgerm}
\usepackage{geometry}
\usepackage{relsize}
\usepackage{afterpage}
\usepackage{lmodern}
\usepackage{tabularx}
\usepackage{cancel}
\usepackage{dutchcal} 
\usepackage{natbib} 

\newcommand{\dd}{\mathrm{d}}

\newcommand{\vae}{\varepsilon}
\newcommand{\vt}{\vartheta}

\newcommand{\trans}{^\mathsf{T}}
\newcommand{\transl}{\mathscr{V}}
\newcommand{\av}{\bm{a}}

\newcommand{\cv}{\bm{c}}

\newcommand{\dv}{\bm{d}}

\newcommand{\e}{\bm{e}}
\newcommand{\x}{\bm{x}}
\newcommand{\zero}{\bm{0}}
\newcommand{\y}{\bm{y}}

\newcommand{\normal}{\bm{\nu}}
\newcommand{\surface}{\mathscr{S}}
\newcommand{\tplane}{\mathscr{T}}
\newcommand{\free}{\mathscr{F}}
\newcommand{\nablas}{\nabla\!_\mathrm{s}}
\newcommand{\nablast}{\nabla^\ast\!\!\!_\mathrm{s}}

\newcommand{\uv}{\bm{u}}

\newcommand{\h}{\bm{h}}

\newcommand{\bv}{\bm{b}}
\newcommand{\wv}{\bm{w}}

\newcommand{\vv}{\bm{v}}

\newcommand{\W}{\mathbf{W}}
\newcommand{\F}{\mathbf{F}}
\newcommand{\G}{\mathbf{G}}
\newcommand{\Gd}{\mathbf{G}_\mathrm{d}}
\newcommand{\Gb}{\mathbf{G}_\mathrm{b}}
\newcommand{\I}{\mathbf{I}}

\newcommand{\R}{\mathbf{R}}
\newcommand{\Rd}{\mathbf{R}_\mathrm{d}}
\newcommand{\Rb}{\mathbf{R}_\mathrm{b}}
\newcommand{\Q}{\mathbf{Q}}

\newcommand{\Rr}{\R'_\vae}

\newcommand{\C}{\mathbf{C}}
\newcommand{\B}{\mathbf{B}}
\newcommand{\U}{\mathbf{U}}
\newcommand{\V}{\mathbf{V}}

\newcommand{\tr}{\operatorname{tr}}
\newcommand{\co}{\operatorname{co}}

\newcommand{\curvature}{(\nablas\normal)}

\newcommand{\A}{\mathbf{A}}

\newcommand{\euclid}{\mathscr{E}}

\newcommand{\framen}{(\e_1,\e_2,\normal)}
\newcommand{\framee}{(\e_1,\e_2,\e_3)}

\newcommand{\tangent}{\bm{t}}
\newcommand{\curve}{\bm{c}}

\newcommand{\orth}{\mathsf{SO}(3)}
\newcommand{\lin}{\mathsf{GL}(3)}
\newcommand{\Lin}[1]{{\mathsf{GL}({#1})}}

\newcommand{\proj}{\mathbf{P}(\normal)}

\newcommand{\Proj}{\mathbf{P}}

\newcommand{\sphere}{\mathbb{S}^2}
\newcommand{\disc}{\mathbb{S}^1}
\newcommand{\Disc}{\mathbb{D}_{\ell}}

\newcommand{\arccosh}{\operatorname{arccosh}}
\newcommand{\ave}[1]{{\left\langle{#1}\right\rangle}}

\newcommand{\ws}{w_\mathrm{s}}
\newcommand{\wdr}{w_\mathrm{d}}
\newcommand{\wb}{w_\mathrm{b}}

\theoremstyle{definition}

\begin{document}
	\latintext
	\title{A Variational Theory for Soft Shells}
	\author{Andr\'e M. Sonnet}
	\email{Andre.Sonnet@strath.ac.uk}
	\affiliation{Department of Mathematics and Statistics, University of Strathclyde, 26 Richmond Street, Glasgow, G1 1XH, U.K. }
	\author{Epifanio G. Virga}
	\email{eg.virga@unipv.it}
	\affiliation{Dipartimento di Matematica, Universit\`a di Pavia, Via Ferrata 5, 27100 Pavia, Italy }

	\date{\today}

	\begin{abstract}
	Three general modes are distinguished in the deformation of a thin shell; these are \emph{stretching}, \emph{drilling}, and \emph{bending}. Of these, the drilling mode is the one more likely to emerge in a \emph{soft matter} shell (as compared to a hard, structural one), as it is ignited by a swerve of material fibers about the local normal. We propose a hyperelastic theory for soft shells, based on a separation criterion that envisages the strain-energy density as the sum of three independent pure measures of stretching, drilling, and bending. Each individual measure is prescribed to vanish on all other companion modes. The result is a direct, second-grade theory featuring a bending energy \emph{quartic} in an invariant strain descriptor that stems from the polar rotation hidden in the deformation gradient (although quadratic energies are also appropriate in special cases). The proposed energy functional has a multi-well character, which fosters cases of \emph{soft elasticity} (with a continuum of ground states) related to minimal surfaces.
	\end{abstract}

	\maketitle

\section{Introduction}\label{sec:intro}
The theories for elastic shells are plentiful. They fall into two broad categories: \emph{direct} and \emph{derived}. The former envision a shell as a material surface in three-dimensional space subject to the balance laws of continuum mechanics as fit for a low-dimensional body. The latter, on the other hand, envision a shell as a \emph{thin} three-dimensional body described by the standard balance laws of continuum mechanics, which by the unbalance of sizes can be written in various approximate forms.\footnote{The reader is referred to Sects.~212 and 213 of \cite{truesdell:classical} for a classical summary of these approaches. We also recommend readng Sects.~13--15 of Chapt.~XIV in Antman's book \cite{antman:nonlinear}.}

A bridge between these realms can be established by the theory of the Cosserat brothers \cite{cosserat:theorie,cosserat:theorie_livre},\footnote{As appropriately pointed out in \cite[p.\,389]{truesdell:non-linear},
\begin{quote} 
For continua of one, two, and three dimensions, they developed their theory with elegance, precision, and thoroughness. It seems to have attracted slight attention.
\end{quote}
The last assertion does no longer apply, especially in view of the conspicuous body of recent literature \cite{ghiba:isotropic_I,ghiba:isotropic_II,ghiba:constrained,ghiba:linear_2021,ghiba:linear_2023,ghiba:essay}.
}
whose use of \emph{directors}\footnote{The terminology of \emph{directors} was first introduced in the paper \cite{ericksen:exact} on a general treatment of ordered media.} for low-dimensional bodies can also be regarded as a means to recover information about the lost dimensions, as shown in the classical essay \cite{naghdi:theory}. 

Writing yet another paper proposing a theory for the equilibrium of shells may feel like an \emph{exercise in style}.\footnote{Having in mind Queneau's book \cite{queneau:exercises}, where the same story is retold 99 times in different styles, an endeavour that was also recently emulated in mathematics by the book of Ording \cite{ording:99}, where a single, simple statement is given 99 different proofs (see also the witty review \cite{watkins:99}).} However, rather than merely retelling an old story, here we pursue a line of reasoning that can be traced back to the work of Acharia~\cite{acharya:nonlinear} (and  was further developed in \cite{wood:contrasting,vitral:dilation,virga:pure,acharya:mid-surface}) aiming at identifying \emph{pure} measures of deformations for shells (and plates).\footnote{The comprehensive commentary in \cite{ghiba:essay} is especially recommended in this regard.} Our approach is direct: we shall regard a shell as a smooth material surface $\surface$ in three-dimensional space.

Besides the customary stretching deformation mode, which raises no issue and has its own standard pure measure, by the unique decomposition of the polar rotation\footnote{This is the rotation extracted from the polar decomposition of the deformation gradient.} described in \cite{sonnet:bending-neutral}, we identify \emph{two} further deformation modes, \emph{drilling} and \emph{bending}. The former involves rotations about the local normal to $\surface$ and the latter rotations about an axis on the local tangent plane.\footnote{We could trace back the definition of \emph{drilling} rotation to the paper \cite{hughes:drilling}.} 

A \emph{separation} criterion will guide our quest for pure measures of drilling and bending: for each deformation mode, we shall identify an invariant scalar that also vanishes on all local realizations of the companion mode. To accomplish this task, we start from local averages of two frame-dependent kinematic contents. The averaging process, which is reminiscent of the way order parameters are defined in soft matter physics, delivers invariant measures depending on the  polar rotation gradient, which makes our theory effectively \emph{second-grade}, although in an indirect way. It is  not too much of a stretch to say that the averages from which pure drilling and bending measures originate are special incarnations of the Cosserat directors.

If anything, perhaps our theory differs the most from other direct theories in the attention paid to the drilling mode, which we reckon should be the more excitable the softer is the material constituting the shell, as drilling rotations would then be less hampered. This intuition makes us believe that our theory should be especially fit for \emph{soft} shells.

The paper is organized as follows. In Sect.~\ref{sec:kinematics}, we collect a number of kinematic preliminaries with the dual intent to make the paper self-contained and to accustom the reader to the notation employed here. Section~\ref{sec:theory} presents the theory: it starts from the definition of the local averages and ends with the pure measures and their geometric interpretations, which aid to give the proposed strain-energy density a form more directly connected with both stretching and curvature tensors. Finally, in Sect.~\ref{sec:conclusion}, we summarize the conclusions of our study and reflect on its possible further evolution. The paper is closed by five appendices with a number of ancillary results. In particular, Appendix~\ref{sec:soft_elasticity} illustrates explicit cases of \emph{soft elasticity}, where a class of deformations inhabits the shell's ground state as a consequence of the \emph{multiple} wells present in the strain energy.

\section{Kinematics}\label{sec:kinematics}
In our theory, a \emph{soft} shell is represented as a \emph{material surface} with reference configuration $\surface$, which is assumed to be a \emph{smooth} orientable surface in three-dimensional Euclidean space $\euclid$ endowed with a translation space $\transl$.\footnote{In particular, here $\surface$ will be assumed to be at least of class $C^2$.}
We shall use an intrinsic approach to surface calculus, generally inspired by the work of Weatherburn~\cite{weatherburn:differential_1,weatherburn:differential_2}, whose essential features are also outlined in \cite{sonnet:bending-neutral}. We refer the reader to Appendix~\ref{sec:surface_calculus} for a brief summary of useful definitions and properties that should make our development self-contained.

\begin{figure}[h]
	\centering\includegraphics[width=.4\linewidth]{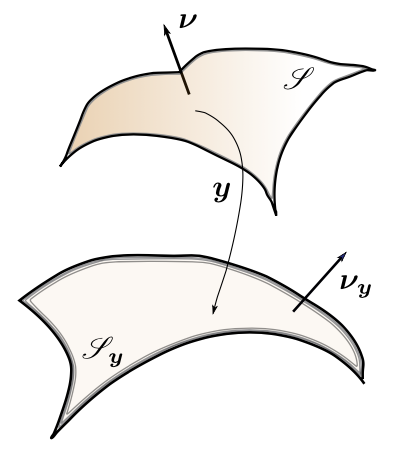}
	\caption{In its reference configuration, a shell is represented by a smooth surface $\surface$ oriented by the unit normal $\normal$. The deformation $\y$ brings $\surface$ into the surface $\surface_{\y}=\y(\surface)$ oriented by the unit normal $\normal_{\y}$.}
	\label{fig:sketches}
\end{figure}

As shown in Fig.~\ref{fig:sketches},
a deformed configuration $\surface_{\y}$ of the shell $\surface$ is represented by a mapping $\y:\surface\to\euclid$, which we assume twice-differentiable, so that $\surface_{\y}=\y(\surface)$. Extending to the present setting results well-known from three-dimensional kinematics (see, for example, Chapt.~6 of \cite{gurtin:mechanics}), we can represent the deformation gradient $\nablas\y$ as
\begin{equation}
	\label{eq:deformation_gradient}
	\nablas\y=\lambda_1\vv_1\otimes\uv_1+\lambda_2\vv_2\otimes\uv_2,
\end{equation}
where the positive scalars $\lambda_1$, $\lambda_2$ are the \emph{principal stretches}, while $\uv_1$, $\uv_2$ and $\vv_1$, $\vv_2$ are the \emph{right} and \emph{left} principal \emph{directions of stretching}, characterized as the (normalized) eigenvectors of the stretching tensors
\begin{subequations}\label{eq:stretching_tensors}
\begin{eqnarray}
\U:=&\sqrt{\C}=\lambda_1\uv_1\otimes\uv_1+\lambda_2\uv_2\otimes\uv_2,\label{eq:stretching_tensor_U}\\
\V:=&\sqrt{\B}=\lambda_1\vv_1\otimes\vv_1+\lambda_2\vv_2\otimes\vv_2,\label{eq:stretching_tensor_V}
\end{eqnarray}
\end{subequations}
where
\begin{equation}
	\label{eq:C_B_tensors}
	\C:=(\nablas\y)\trans(\nablas\y)\quad\text{and}\quad\B:=(\nablas\y)(\nablas\y)\trans
\end{equation}
are the \emph{right} and \emph{left} Cauchy-Green tensors, respectively. It should be noted that, for a point on $\surface$ represented by the position vector $\x$, $\U(\x)$ is a linear mapping of $\tplane_{\x}$ onto itself, that is, $\U(\x)\in\Lin{\tplane_{\x}}$, whereas $\V(\y(\x))$ is a linear mapping of $\tplane_{\y(\x)}$ onto itself, that is, $\V(\y(\x))\in\Lin{\tplane_{\y(\x)}}$.\footnote{Here $\tplane_{\x}$ is the plane tangent to $\surface$ at $\x$ and $\tplane_{\y(\x)}$ is the plane tangent to $\surface_{\y}$ at $\y(\x)$, see also Appendix~\ref{sec:surface_calculus}.}

By the polar decomposition theorem for the deformation of surfaces proved in \cite{man:coordinate} in the spirit of the general coordinate-free theory of material surfaces of Gurtin and Murdoch~\cite{gurtin:continuum,gurtin:addenda}, which we also embrace, the deformation gradient $\nablas\y$ can also be written as
\begin{equation}
	\label{eq:polar_decomposition}
	\nablas\y=\R\U=\V\R,
\end{equation}
where the rotation $\R$ is an element  of the special orthogonal group $\orth$ in three-dimensional translation space $\transl$; it will also be called the \emph{polar} rotation. It readily follows from \eqref{eq:deformation_gradient} and \eqref{eq:polar_decomposition} that $\vv_i=\R\uv_i$ for $i=1,2$.

As also recalled in \cite{sonnet:bending-neutral}, a rotation $\R\in\orth$ can be represented in terms of a vector $\av\in\transl$ through Rodrigues' formula \cite{altmann:hamilton},
\begin{equation}
	\label{eq:rodrigues_formula}
	\R(\av)=\frac{1}{1+a^2}\{(1-a^2)\I+2\av\otimes\av+2\W(\av)\},
\end{equation}
where $a^2:=\av\cdot\av$, $\I$ is the  second-rank identity tensor, and $\W(\av)$ is the skew-symmetric tensor associated with $\av$.\footnote{The action of $\W(\av)$ on any vector $\vv\in\transl$ is given by $\W(\av)\vv=\av\times\vv$, where $\times$ denotes the vector product.} The connection between \eqref{eq:rodrigues_formula} and  Euler's classical representation,
\begin{equation}
	\label{eq:euler_formula}
	\R=\I+\sin\alpha\W(\e)+(1-\cos\alpha)\W(\e)^2,
\end{equation}
where $\e\in\sphere$ is a unit vector designating the \emph{axis} of rotation and $\alpha\in[0,\pi]$ is the \emph{angle} of rotation, is established by setting
\begin{equation}
	\label{eq:a_vector}
	\av=\tan\left(\frac\alpha2\right)\e.
\end{equation}
Thus, a $\pi$-turn, which is represented by \eqref{eq:euler_formula} for $\alpha=\pi$, is  represented by \eqref{eq:rodrigues_formula} for $a\to\infty$. Equation \eqref{eq:rodrigues_formula} can easily be inverted, as the vector $\av$ representing $\R$ is identified by 
\begin{equation}
	\label{eq:W(a)}
	\W(\av)=\frac{\R-\R\trans}{1+\tr\R},
\end{equation}
from which it follows that 
\begin{equation}
	\label{eq:a^2}
	a^2=\frac{3-\tr\R}{1+\tr\R},
\end{equation}
which shows that all $\pi$-turns are characterized by having $\tr\R=-1$.\footnote{Equation \eqref{eq:a^2} is a consequence of the identity $\tr\R^2=(\tr\R)^2-2\tr\R$, valid for all $\R\in\orth$; the latter follows from the Cayley-Hamilton theorem applied to $\R$.}

By the decomposition formula of Rodrigues \cite{rodrigues:lois}, for every rotation $\R(\av)$ and a given direction in space there is a single pair of vectors $(\av_1,\av_2)$, with $\av_1$ along the given direction and $\av_2$ orthogonal to it, such that $\R(\av)=\R(\av_2)\R(\av_1)$ (see \cite{sonnet:bending-neutral}, where a variational interpretation of this decomposition is also provided). Thus, at any point $\x\in\surface$, choosing $\av_1$ along $\normal$, we can decompose the rotation $\R$ in \eqref{eq:polar_decomposition} as 
\begin{equation}
	\label{eg:rotation_decomposition}
	\R(\av)=\R(\bv)\R(\dv)
\end{equation}
where 
\begin{equation}
	\label{eq:d_and_b_vectors}
	\dv=a_\nu\normal,\quad \quad\bv=\frac{1}{1+a_\nu^2}\{\I+a_\nu\W(\normal)\}\proj\av,\quad\text{with}\quad a_\nu:=\av\cdot\normal,
\end{equation}
and $\proj:=\I-\normal\otimes\normal$ is the projection onto the tangent plane. We call $\R(\dv)$ the \emph{drilling} rotation and $\R(\bv)$ the \emph{bending} rotation.

We see from \eqref{eq:d_and_b_vectors} that whenever $\R(\av)$ is a $\pi$-turn, so is also $\R(\dv)$, unless $a_\nu=0$, in which case $\dv=\bm{0}$ and $\bv=\av$. For a given deformation $\y$, both $\dv$ and $\bv$ are vector fields on $\surface$; we shall call them drilling and bending \emph{contents} of $\y$, respectively. Each of them can be associated with a rotation angle, denoted $\alpha_\mathrm{d}$ and $\alpha_\mathrm{b}$, respectively, which in accordance with \eqref{eq:a_vector} are defined as
\begin{equation}
	\label{eq:drilling_bending_angles}
	\alpha_\mathrm{d}:=2\arctan d\quad\text{and}\quad\alpha_\mathrm{b}:=2\arctan b,
\end{equation}
where $d:=\sqrt{\dv\cdot\dv}$ and $b:=\sqrt{\bv\cdot\bv}$. We call the angles $\alpha_\mathrm{d}$ and $\alpha_\mathrm{b}$  the \emph{drilling} and \emph{bending} angles, respectively.

The vectors $\dv$ and $\bv$ constitute the \emph{non-metric} kinematic contents of the deformation $\y$  in a given frame, while the stretching tensors $\U$ and $\V=\R\U\R\trans$ constitute its \emph{metric} kinematic contents (in the same frame). Thus, an \emph{isometric} deformation occurs whenever $\U=\proj$, and correspondingly $\V=\Proj(\normal_{\y})$, where $\normal_{\y}$ is the unit normal that orients $\surface_{\y}=\y(\surface)$. A deformation $\y$ for which there is a frame where $\bv(\x)=\zero$ at a point $\x\in\surface$ is locally a \emph{pure drilling}; it may also be called a \emph{bending-neutral} deformation because when composed with another deformation it leaves the bending content of the latter unchanged (a formal proof of this property is given in Appendix~\ref{sec:bending_neutral}). More restrictively, a deformation $\y$ for which there is a frame where $\dv(\x)=\zero$ is a \emph{pure bending}, if the direction of $\bv(\x)$ does not swerve (at least locally) on the tangent plane of $\surface$.\footnote{This requirement is enforced by setting $\nablas\hat{\bv}(\x)=\zero$ for the unit vector $\hat{\bv}$ of $\bv$. We shall discuss in Appendix~\ref{sec:soft_elasticity} the consequences of a less restrictive definition of pure bending.}

\begin{figure}[h]
	\centering\includegraphics[width=.8\linewidth]{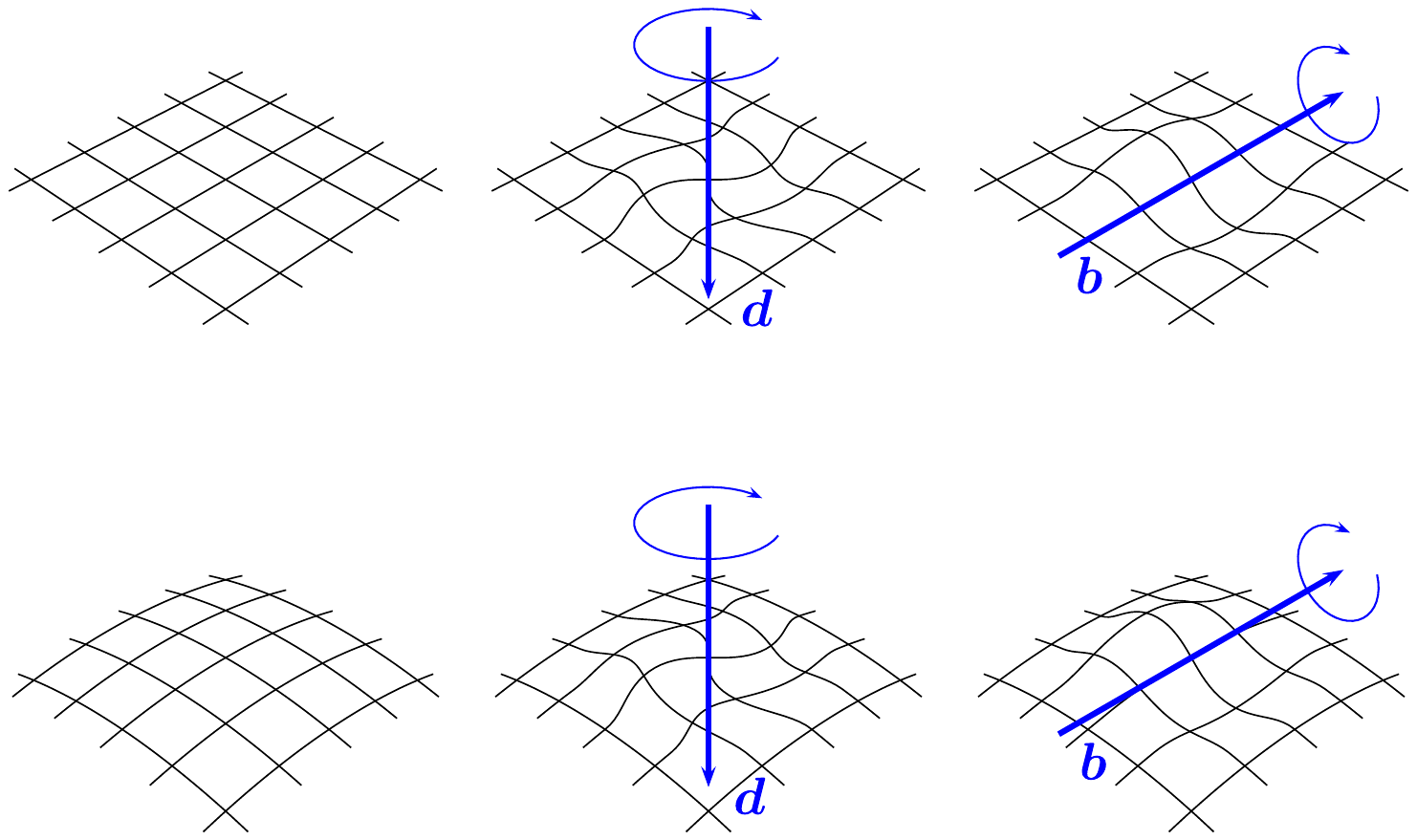}
	\caption{Pictorial representation of local pure drilling and bending deformations. The upper and lower rows show how a flat and a curved surface are correspondingly affected by the two independent deformation modes. The drilling and bending contents, $\dv$ and $\bv$, are represented alongside the corresponding drilling and bending angles, $\alpha_\mathrm{d}$ and $\alpha_\mathrm{b}$, defined by \eqref{eq:drilling_bending_angles}.}
	\label{fig:drill_and_bend}
\end{figure}

Figure~\ref{fig:drill_and_bend} represents pictorially how a pure drilling and a pure bending may affect locally a surface originally flat or curved.

\emph{Pure} measures of bending were defined in \cite{virga:pure} as strain measures  invariant under all bending-neutral deformations. It was proved in \cite{sonnet:bending-neutral} that bending-neutral deformations actually exist in the large. They preserve \emph{minimal surfaces}, in the sense that a minimal surface is mapped into another minimal surface by a bending-neutral deformation, and for any given pair of such surfaces with the same Gauss map,\footnote{That is, which have one and the same system of normals (possibly, to within a uniform rigid rotation).} there is a bending-neutral deformation mapping one into the other.

The positive scalars $d^2$ and $b^2$ (along with the corresponding angles $\alpha_\mathrm{d}$ and $\alpha_\mathrm{b}$) will play a special role in our development. They are related to the vector $\av$ that represents $\R$ through the equations (see also \cite{sonnet:bending-neutral})
\begin{equation}
	\label{eq:d_and_b_scalars}
	d^2=a_\nu^2\quad\text{and}\quad b^2=\frac{a^2-a_\nu^2}{1+a_\nu^2}.
\end{equation}
By use of \eqref{eq:W(a)} and \eqref{eq:a^2}, these can also be given the following form\footnote{Here the inner product of two second-rank tensors, $\A$ and $\B$, is defined as $\A\cdot\B:=\tr(\A\trans\B)=\tr(\B\trans\A)$.}
\begin{equation}
	\label{eq:d_and_b_scalars_explicit}
	d^2=\left(\frac{\R\cdot\W(\normal)}{1+\tr\R}\right)^2\quad\text{and}\quad b^2=\frac{(3-\tr\R)(1+\tr\R)-(\R\cdot\W(\normal))^2}{(1+\tr\R)^2+(\R\cdot\W(\normal))^2}.
\end{equation}

To relate both $d^2$ and $b^2$ directly to the deformation gradient $\nablas\y$, we need to extract the polar rotation $\R$ from the latter. To this end, we remark that $\R$ can be represented as follows,
\begin{equation}
	\label{eq:R_representation}
	\R=\vv_1\otimes\uv_1+\vv_2\otimes\uv_2+\normal_{\y}\otimes\normal
\end{equation}
and that $\normal_{\y}=\vv_1\times\vv_2$, if we have oriented $\surface$ according to the choice $\normal=\uv_1\times\uv_2$. Since, by \eqref{eq:deformation_gradient} and \eqref{eq:stretching_tensors},
\begin{equation}
\label{eq:rotation_pre_extraction}
(\nablas\y)\U^{-1}=\vv_1\otimes\uv_1+\vv_2\otimes\uv_2,
\end{equation}
we have that
\begin{equation}
	\label{eq:normal_ast}
	\normal_{\y}=(\nablas\y)\U^{-1}\uv_1\times(\nablas\y)\U^{-1}\uv_2=\co((\nablas\y)\U^{-1})\normal,
\end{equation}
where $\co(\cdot)$ denotes the cofactor of a tensor (as defined, for example, in Sect.~2.11 of \cite{gurtin:mechanics}). By combining \eqref{eq:R_representation}, \eqref{eq:rotation_pre_extraction}, and \eqref{eq:normal_ast}, we arrive at 
\begin{equation}
	\label{eq:rotation_extraction}
	\R=(\nablas\y)\U^{-1}+\co((\nablas\y)\U^{-1}),
\end{equation}
where we have extended the tensor $(\nablas\y)\U^{-1}$ to be an element of $\lin$ by requiring that $(\nablas\y)\U^{-1}\normal=\zero$.\footnote{For such an extension, it is a simple matter to prove that $\co((\nablas\y)\U^{-1})\uv_i=\zero$, for $i=1,2$.} 

The non-metric kinematic contents, $\dv$ and $\bv$, of the deformation $\y$ express  independent degrees of freedom coexisting in the polar rotation $\R$. While the bending content $\bv$ is expected to be significant in the deformation of all shells, irrespective of the material that constitutes them, the drilling content $\dv$ is expected to be significant especially in the deformation of shells constituted by materials sufficiently 
\emph{soft} to allow for surface fibers to be wrung in the tangent plane.

Following \cite[Chapt.\,20]{gurtin:mechanics} we consider the surface $\surface$ as living in a \emph{reference} space of its own (also called the \emph{material} space), thus remaining unaffected by a change of frame that transforms the deformation $\y$ into
\begin{equation}
	\label{eq:y_ast}
	\y^\ast(\x)=\y_0(t)+\Q(t)\y(\x),
\end{equation}
where $\y_0(t)\in\transl$ is a translation and $\Q(t)\in\orth$ is a rotation, both possibly depending on time $t$.

Under the action of a change of frame, the principal directions $\uv_i$ remain unchanged, as do all material vectors, whereas the principal directions $\vv_i$ are transformed into $\vv_i^\ast=\Q\vv_i$. Thus, the stretching tensor $\U$ is \emph{frame-invariant}, as $\U^\ast=\U$, while the stretching tensor $\V$ is \emph{frame-indifferent}, as $\V^\ast=\Q\V\Q\trans$.

On the other hand, the polar rotation $\R$ transforms into $\R^\ast=\Q\R$, and so it is neither frame-invariant nor frame-indifferent.  Accordingly, it follows from \eqref{eq:W(a)} and \eqref{eq:a^2} that
\begin{equation}
	\label{eq:change_frame_W(a)}
	\W(\av^\ast)=\frac{\Q\R-\R\trans\Q\trans}{1+\tr(\Q\R)}\quad\text{and}\quad {a^\ast}^2=\frac{3-\tr(\Q\R)}{1+\tr(\Q\R)},
\end{equation}
which are not neatly related to $\W(\av)$ and $a^2$. Even more intricate transformation laws would follow from \eqref{eq:d_and_b_scalars_explicit} for both $d^2$ and $b^2$.

Despite a naive intuition which would suggest to take $d^2$ and $b^2$ as scalar \emph{measures} of the two independent components of $\R$, neither of them could be one, as they are frame-dependent. We find it instructive to illustrate in Appendix~\ref{sec:contents_not_measures} how complicated  this dependence could be, even in a simple, innocent-looking case.

In the following section, we shall extract from $d^2$ and $b^2$ appropriate frame-invariant measures of drilling and bending.

\section{Theory}\label{sec:theory}
In this section, we attempt to build a theory for shells based upon an elastic stored energy where the stretching degrees of freedom of the material surface $\surface$ representing the mechanical system are combined on equal footing with the \emph{non-stretching} degrees of freedom. Among these, our kinematic analysis has distinguished two independent components, that is, drilling and bending, the former perhaps more distinctive of soft shells than the latter.

It is quite clear how the stretching of $\surface$ can be measured for a material (transversely) isotropic about the normal $\normal$: a natural pure stretching measure $w_\mathrm{s}$ is the mismatch between the (two-dimensional) stretching tensor $\U$ and the projection $\proj$, which would correspond to an isometric deformation,\footnote{Here, for a second-rank tensor $\A$, $|\A|^2:=\A\cdot\A$. Moreover, as customary in Koiter's model \cite{koiter:consistent}, one could also set
$$
w_\mathrm{s}=|\C-\proj|^2=|\B-\Proj(\normal_{\y})|^2,
$$
which would make $\ws$  \emph{quartic} in the principal stretches. However, stretching measures are not our primary interest in this paper.}
\begin{equation}
	\label{eq:w_s}
	w_\mathrm{s}=|\U-\proj|^2=|\V-\Proj(\normal_{\y})|^2.
\end{equation}

To define similar, appropriate pure measures of \emph{drilling} and \emph{bending} $w_\mathrm{d}$ and $w_\mathrm{b}$, we start from $d^2$ and $b^2$, respectively. We need to infer from the latter a content unaffected by a uniform rotation: this is achieved by taking \emph{local} averages over \emph{relative} rotations, as illustrated below.

\subsection{Relative rotation local averaging}\label{sec:average}
To suppress any uniform background rotation, which is the source of frame-dependence of $d^2$ and $b^2$, we take (transversely isotropic) averages over relative rotations in a small disk on the tangent plane. We thus obtain intrinsic measures of drilling and bending, at least for materials  isotropic about the  shell's normal. The origin of these measures is formally akin to the definition of scalar \emph{order parameters} in soft matter physics; they play in our theory a role similar to that played by \emph{directors} in the classical Cosserat theory.

For $\x\in\surface$, let $\Disc$ be a (small) disk of radius $\ell$ and centre  on the tangent plane $\tplane_{\x}$. The points $\x_\varepsilon$ of $\Disc$ are described by
\begin{equation}
	\label{eq:x_epsilon}
	\x_\varepsilon=\x+\varepsilon\uv,
\end{equation}
where $\varepsilon$ ranges in the interval $[0,\ell]$ and $\uv\in\disc$ is a unit vector on $\tplane_{\x}$. By lifting onto $\Disc$ along the normal $\normal$ at $\x$ the neighbourhood of $\x$ intercepted on $\surface$ by the cylinder based on $\Disc$ (following the method introduced in \cite{gurtin:continuum}),\footnote{An alternative method, introduced in \cite{arroyo:atomistic}, makes use of the \emph{exponential map} for surfaces (see also p.\,77 of \cite{morgan:riemannian}.)} we can define on $\Disc$ both fields $\R_\varepsilon:=\R(\x_\varepsilon)$ and $\normal_\varepsilon:=\normal(\x_\varepsilon)$.

For given $\uv\in\disc$ such that $\normal\cdot\uv=0$, we expand both $\R_\varepsilon$ and $\normal_\varepsilon$ up to the lowest order in $\varepsilon$,
\begin{subequations}
	\begin{align}
		\R_\varepsilon&=\R+\varepsilon(\nablas\R)\uv+o(\varepsilon),\label{eq:R_epsilon}\\
		\normal_\varepsilon&=\normal+\varepsilon(\nablas\normal)\uv+o(\varepsilon),\label{eq:normal_epsilon}
\end{align}	
\end{subequations}
where $(\nablas\R)\uv\in\lin$ is a second-rank tensor (obtained by contracting the third leg of $\nablas\R$ with $\uv$). It readily follows from \eqref{eq:R_epsilon} that the \emph{relative} rotation $\R'_\varepsilon$ is represented on $\Disc$ as
\begin{subequations}\label{eq:R_prime_normal_epsilon}
\begin{equation}
	\label{eq:R_prime_epsilon}
	\R'_\varepsilon:=\R\trans\R_\varepsilon=\I+\varepsilon\W_{\uv}+o(\vae),
\end{equation}
where $\W_{\uv}:=\R\trans(\nablas\R)\uv$ is a skew-symmetric tensor (depending linearly on $\uv$), and from \eqref{eq:normal_epsilon} it follows that 
\begin{equation}
	\label{eq:W_normal_epsilon}
	\W(\normal_\vae)=\W(\normal)+\vae\W(\curvature\uv)+o(\vae).
\end{equation}
\end{subequations}

With the aid of \eqref{eq:d_and_b_scalars_explicit}, the \emph{relative average} drilling content is then defined as
\begin{equation}
	\label{eq:averaged_drilling_definition}
	\ave{d^2}:=\frac{1}{A(\Disc)}\int_0^\ell\vae\dd\vae\int_{\disc}\left(\frac{\R'_\vae\cdot\W(\normal_\vae)}{1+\tr\R'_\vae} \right)^2\dd\uv,
\end{equation}
where $A$ denotes the area measure. By use of equations \eqref{eq:R_prime_normal_epsilon},  we also give \eqref{eq:averaged_drilling_definition} the form
\begin{align}
	\ave{d^2}&=\frac{\ell^2}{32}\ave{(\W_{\uv}\cdot\W(\normal))^2}_{\uv\in\disc}+o(\ell^2)\nonumber\\
	&=\frac{\ell^2}{64}|\W(\normal)\circ\G|^2+o(\ell^2),	\label{eq:averaged_drilling}
\end{align}
where we have set
\begin{equation}
	\label{eq:G_definition}
	\G:=\R\trans\nablas\R,
\end{equation}
a third-rank tensor,\footnote{$\G$ is somewhat reminiscent of the third-rank tensor $\F\trans\nablas\F$, where $\F:=\nablas\y$, that Murdoch~\cite{murdoch:direct} employed for his direct second-grade hyperelastic theory of shells and that has recently received a full kinematic characterization \cite{tiwari:characterization}.} which in a generic Cartesian frame $\framee$ would have components
\begin{equation}
	\label{eq:G_components}
	G_{ijk}=R_{hi}R_{hj;k},
\end{equation}
if $R_{hi}$ and $R_{hj;k}$ denote the components (in the same frame) of $\R$ and $\nablas\R$, respectively.\footnote{Here and below, we shall use the usual convention of summing over repeated indices. Moreover, as also recalled in Appendix~\ref{sec:surface_calculus}, a semicolon denotes surface differentiation.} In \eqref{eq:averaged_drilling_definition}, $\W(\normal)\circ\G$ is a vector of $\transl$, as for a generic second-rank tensor $\A$ and any triadic component $\av\otimes\bv\otimes\cv$ of $\G$, 
\begin{equation}
	\label{eq:generic_triadic}
	\A\circ\G:=(\av\cdot\A\bv)\cv.
\end{equation}

More generally, it follows from \eqref{eq:G_definition} that, for any orthonormal basis $(\e_1,
\e_2)$ on the tangent plane $\tplane_{\x}$ to $\surface$, $\G(\x)$ can be represented as
\begin{equation}
	\label{eq:G_representation}
	\G(\x)=\W_1\otimes\e_1+\W_2\otimes\e_2,
\end{equation}
where both $\W_1$ and $\W_2$ are skew-symmetric tensors. Thus, combining \eqref{eq:generic_triadic} and \eqref{eq:G_representation}, we can also write
\begin{equation}
	\label{eq:W_G}
	\W(\normal)\circ\G=(\W_1\cdot\W(\normal))\e_1+(\W_2\cdot\W(\normal))\e_2=2((\wv_1\cdot\normal)\e_1+(\wv_2\cdot\normal)\e_2),
\end{equation}
where $\wv_1$ and $\wv_2$ are the axial vectors of $\W_1$ and $\W_2$, respectively. Equivalently, by introducing a movable orthonormal frame $(\e_1,\e_2,\normal)$ gliding on $\surface$, $\G$ can also be represented in the form
\begin{equation}
	\label{eq:G_representation_alternative}
	\G=\W(\e_1)\otimes\av_1+\W(\e_2)\otimes\av_2+\W(\normal)\otimes\av_3,
\end{equation}
where $\W(\e_1)$ and $\W(\e_2)$ are the skew-symmetric tensors associated with $\e_1$ and $\e_2$, respectively, and $\av_1$, $\av_2$, and $\av_3$ are vectors such that $\av_i\cdot\normal=0$.

It readily follows from \eqref{eq:W_G} that
\begin{equation}
	\label{eq:averaged_drilling_rewritten}
	|\W(\normal)\circ\G|^2=4((\wv_1\cdot\normal)^2+(\wv_2\cdot\normal)^2).
\end{equation}

Similarly, again by \eqref{eq:d_and_b_scalars_explicit}, we arrive at the following expression for the \emph{relative average} bending content,
\begin{align}
	\label{eq:averaged_bending}
	\ave{b^2}&:=\frac{1}{A(\Disc)}\int_0^\ell\vae\dd\vae\int_{\disc}\frac{(3-\tr\Rr)(1+\tr\Rr)-(\Rr\cdot\W(\normal_\vae))^2}{(1+\tr\Rr)^2+(\Rr\cdot\W(\normal_\vae))^2}\dd\uv,\nonumber\\
	&=\frac{\ell^2}{32}\ave{2\W_{\uv}\cdot\W_{\uv}-(\W_{\uv}\cdot\W(\normal))^2}_{\uv\in\disc}+o(\ell^2),\nonumber\\
	&=\frac{\ell^2}{32}\left\{|\G|^2-\frac12|\W(\normal)\circ\G|^2\right\}+o(\ell^2),
\end{align}
where, by \eqref{eq:G_representation}, 
\begin{equation}
	\label{eq:G_squared}
	|\G|^2=|\W_1|^2+|\W_2|^2=2(w_1^2+w_2^2).
\end{equation}

The detailed computations that prove \eqref{eq:averaged_drilling} and \eqref{eq:averaged_bending} are an easy exercise presented in Appendix~\ref{sec:surface_calculus}; here we only note that they make use of the identity
\begin{equation}
	\label{eq:average_identity}
	\ave{\uv\otimes\uv}_{\uv\in\disc}=\frac12\proj,
\end{equation}
which embodies our notion of transverse isotropy.

Since in a change of frame $\R^\ast=\Q\R$ and $\nablas\R=\Q\nablas\R$, it is a simple matter to see from \eqref{eq:G_definition} that $\G^\ast=\G$, and so both $\ave{d^2}$ and $\ave{b^2}$ are frame-invariant, as expected. Moreover, depending only on $\R$, they are plainly invariant under a superimposed \emph{pure extension}, in which $\U$ is replaced by $\lambda\U$  with $\lambda>0$ (and $\V$ by $\lambda\V$). Finally, they both vanish if $\nablas\R$ does, that is, if $\R$ is locally \emph{uniform}.

The scene is now set to extract from $\ave{d^2}$ and $\ave{b^2}$ pure measures of drilling and bending, $\wdr$ and $\wb$, akin to the natural measure of stretching $\ws$ in \eqref{eq:w_s}. Precisely as $\ws$ vanishes on isometries, we wish $\wdr$ and $\wb$ to vanish on pure bending and drilling deformation modes, respectively.\footnote{In addition to vanish on uniform rotations.} We start by building $\wb$.

\subsection{Pure Bending Measure}\label{sec:pure_bending}
A pure drilling mode is embodied by a bending-neutral deformation, for which $\R=\Rd$, where $\Rd$ is a rotation about $\normal$. Here we shall not be concerned with the existence in the large of a bending-neutral deformation; it will suffice to assume that the algebraic condition that makes it compatible with the surface is met locally (see \cite{virga:pure,sonnet:bending-neutral}).
A pure measure of bending $\wb$ is thus constructed from $\ave{b^2}$ (at the leading order in $\ell$ stripped of its scaling factor) by requiring that it vanishes locally on a generic bending-neutral deformation of $\surface$.

To this end, we recall that a movable frame $\framen$ glides over $\surface$ according to the laws
\begin{equation}\label{eq:gliding_laws}
	\begin{cases}
		\nablas\e_1=\e_2\otimes\cv+\normal\otimes\dv_1,\\
		\nablas\e_2=-\e_1\otimes\cv+\normal\otimes\dv_2,\\
		\nablas\normal=-\e_1\otimes\dv_1-\e_2\otimes\dv_2,
	\end{cases}
\end{equation}
where the vector fields $(\cv,\dv_1,\dv_2)$ are everywhere tangent to $\surface$; these are the \emph{connectors} of the movable frame: more precisely, $\cv$ is the \emph{spin} connector and $\dv_1$, $\dv_2$ are the \emph{curvature} connectors. Since the curvature tensor $\nablas\normal$ is symmetric, the curvature connectors must obey the identity
\begin{equation}
	\label{eq:connectors_identity}
	\dv_1\cdot\e_2=\dv_2\cdot\e_1.
\end{equation}
By \eqref{eq:euler_formula}, $\Rd$ can also be represented in the frame $\framen$ in the form
\begin{equation}
	\label{eq:R_d}
	\Rd=\cos\alpha(\e_1\otimes\e_1+\e_2\otimes\e_2)+\sin\alpha(\e_2\otimes\e_1-\e_1\otimes\e_2)+\normal\otimes\normal,
\end{equation}
where $\alpha$ is a scalar field.

Repeated use of \eqref{eq:gliding_laws} in \eqref{eq:R_d} leads us to express $\Gd:=\Rd\trans\nablas\Rd$ as
\begin{equation}
	\label{eq:G_d}
	\Gd=-\W(\e_1)\otimes((1-\cos\alpha)\dv_2+\sin\alpha\dv_1)+\W(\e_2)\otimes((1-\cos\alpha)\dv_1-\sin\alpha\dv_2)+\W(\normal)\otimes\nablas\alpha,
\end{equation}
which specializes \eqref{eq:G_representation_alternative}. By direct computation, we then obtain that 
\begin{align}
	\label{eq:pre_w_b}
	|\Gd|^2-\frac12|\W(\normal)\circ\Gd|^2&=4(1-\cos\alpha)(d_1^2+d_2^2)\nonumber\\&=4(1-\cos\alpha)|\nablas\normal|^2=4\normal\cdot\Gd\circ\nablas\normal,
\end{align}
where $\Gd\circ\nablas\normal$ is a vector defined according to the convention that for a generic triadic component $\av\otimes\bv\otimes\cv$ of $\G$ and a generic second-rank tensor $\A$ sets
\begin{equation}
	\label{eq:G_A}
	\G\circ\A:=(\bv\cdot\A\cv)\av.
\end{equation}

Thus the simplest invariant pure (positive) measure of bending can be defined as\footnote{It is a simple matter to show that failing to square the right hand side of \eqref{eq:w_b} would fail to make $\wb$ positive definite.} 
\begin{equation}
	\label{eq:w_b}
	\wb:=\left(|\G|^2-\frac12|\W(\normal)\circ\G|^2-4\normal\cdot\G\circ\nablas\normal\right)^2.
\end{equation}
This measure is \emph{quartic} in $\G$ and it vanishes for both $\G=\zero$ and $\G=\Gd$. We may say that $\wb$ as defined by \eqref{eq:w_b} has turned $\ave{b^2}$ from a measure of bending into a pure measure of bending.

\subsection{Pure Drilling Measure}\label{sec:pure_drilling}
We proceed in a similar way to extract from $\ave{d^2}$ a pure measure of drilling. In analogy with \eqref{eq:R_d}, letting $\e_2$ be the direction of bending in a local pure bending deformation,\footnote{Apart from the case studied in Appendix~\ref{sec:cylindrical_shells}, pure bending deformations are not known to exist in the large. This is why our considerations here are purely local.} we write the rotation component  $\Rb$ of the deformation gradient as
\begin{equation}
	\label{eq:R_b}
	\Rb=\cos\beta(\e_1\otimes\e_1+\normal\otimes\normal)+\e_2\otimes\e_2+\sin\beta(\normal\otimes\e_1-\e_1\otimes\normal),
\end{equation}
where $\beta$ is a scalar field. Resorting again to \eqref{eq:gliding_laws}, we readily arrive at
\begin{equation}
	\label{eq:G_b}
\begin{split}
	\Gb:=\Rb\trans\nablas\Rb
	&=\W(\e_1)\otimes((\cos\beta-1)\dv_2+\sin\beta\cv)\\&-\W(\e_2)\otimes\nablas\beta+\W(\normal)\otimes((\cos\beta-1)\cv-\sin\beta\dv_2),
\end{split}
\end{equation}
whence it follows that
\begin{equation}
	\label{eq:W_G_b}
	\W(\normal)\circ\Gb=-2((1-\cos\beta)\cv+\sin\beta\dv_2).
\end{equation}
By \eqref{eq:gliding_laws}, requiring that $\nablas\e_2=\zero$ amounts to require that both connectors $\cv$ and $\dv_2$ vanish (possibly only locally), and so also does $\W(\normal)\circ\Gb$. This shows that $\ave{d^2}$ (again at leading order in $\ell$ and stripped of its scaling factor) is itself a pure measure of drilling,
\begin{equation}
	\label{eq:w_d}
	\wdr:=|\W(\normal)\circ\G|^2.
\end{equation}

The reader is referred to Appendix~\ref{sec:soft_elasticity} to see how the definition of $\wdr$ would change when in a pure bending deformation $\nablas\e_2$ is not required to vanish. In Appendix~\ref{sec:cylindrical_shells}, we study instead the case where all admissible deformations are pure bending deformations with $\nablas\e_2$ vanishing everywhere.

\subsection{Energetics}\label{sec:energy}
The pure measures for the different independent degrees of deformation introduced above suggest to define a general elastic stored energy functional in the form
\begin{equation}
	\label{eq:energy_functional}
	\free[\y]:=\int_{\surface}W(\ws,\wdr,\wb)\dd A,
\end{equation}
where $W$ is a positive-definite function of all three measures. The functional $\free$ is properly invariant under a change of frame and keeps separate all independent mechanisms that can deform $\surface$.

The easiest form of $W$ would be the following,
\begin{equation}
	\label{eq:W}
	W=\frac12\mu_\mathrm{s}\ws+\frac12\mu_\mathrm{d}\wdr+\frac14\mu_\mathrm{b}\wb,
\end{equation}
where the positive scalars $\mu_\mathrm{s}$, $\mu_\mathrm{d}$, and $\mu_\mathrm{b}$ are the \emph{stretching}, \emph{drilling}, and \emph{bending} moduli, respectively. While $\ws$ is quadratic in the stretching tensor $\U$ and $\wdr$ is quadratic in $\G$, $\wb$ is quartic in the latter. This is the price to pay to enforce our separation criterion, which requires $\wdr$ and $\wb$ to vanish each on the whole class of deformations that selectively activate the other.

By construction, $W$ is a \emph{multi-well} energy, as it is the sum of potentials with multiple local minima, corresponding to individual local ground states. $W$ can also exhibit cases of \emph{soft elasticity},\footnote{The reader is referred to Chapt.~7 of \cite{warner:liquid} for a general discussion of soft elasticity in nematic elastomers (see also \cite{warner:soft} for an earlier specific contribution).} where it vanishes on a whole class of deformations, as shown in Appendix~\ref{sec:soft_elasticity}, for a class of  deformations generating a family of minimal surfaces.

While $\mu_\mathrm{s}$ in \eqref{eq:W} has the physical dimension of energy per unit area, $\mu_\mathrm{d}$ has that of energy, and $\mu_\mathrm{b}$ of energy times area. Our definition of $\ave{d^2}$ and $\ave{b^2}$ suggest that $\mu_\mathrm{d}$ scales like $\mu_\mathrm{s}\ell^2$, whereas $\mu_\mathrm{b}$ scales like $\mu_\mathrm{s}\ell^4$, where $\ell$ is the size of the averaging disc, totally unrelated to the thickness of the shell (which in our direct theory vanishes by definition).

As shown in Appendix~\ref{sec:cylindrical_shells}, in the presence of an appropriate  constraint on the admissible deformations, also $\wb$ can be given a quadratic form (one that had actually been known since the work of Antman~\cite{antman:general} on elastic rods).

Another notable case where the energy density $W$ need not be quartic in $\G$ is that of plates, for which $\nablas\normal\equiv\zero$, so that $\Gd\circ\nablas\normal$ also vanishes in \eqref{eq:pre_w_b} and our separation criterion can be enforced by simply taking
\begin{equation}
	\label{eq:w_b_plates}
	\widetilde{w}_\mathrm{b}:=|\G|^2-\frac12|\W(\normal)\circ\G|^2
\end{equation}
as a pure bending measure. Thus, $W$ in \eqref{eq:W} can be replaced by
\begin{equation}
	\label{eq:W_plates}
	W=\frac12\mu_\mathrm{s}\ws+\frac12\mu_\mathrm{d}\wdr+\frac12\widetilde{\mu}_\mathrm{b}\widetilde{w}_\mathrm{b},
\end{equation}
which is entirely quadratic. Comparing \eqref{eq:w_d} and \eqref{eq:w_b_plates}, we see that \eqref{eq:W_plates} can be given the following simpler form
\begin{equation}
	\label{eq:W_plates_one_constant}
	W=\frac12\mu_\mathrm{s}\ws+\mu_\mathrm{d}|\G|^2,
\end{equation}
whenever $\widetilde{\mu}_\mathrm{b}=2\mu_\mathrm{d}$, a special instance that defines a useful simplification of the theory.

\subsection{Geometric Representation}\label{sec:representation}
Both $\wdr$ and $\wb$ can be given a more telling representation in terms of mismatches between tensorial quantities with an intrinsic geometric meaning. The formulas that we shall obtain generalize those found in Appendix~\ref{sec:cylindrical_shells} for cylindrical shells and are based on similar calculations, although admittedly more complicated.

We start from the systems of connectors for the movable frames $(\uv_1,\uv_2,\normal)$ and $(\vv_1,\vv_2,\normal_{\y})$ on surfaces $\surface$ and $\surface_{\y}$, respectively. According to \eqref{eq:gliding_laws}, these are the fields $(\cv,\dv_1,\dv_2)$ and $(\cv^\ast,\dv_1^\ast,\dv_2^\ast)$, tangent to $\surface$ and $\surface_{\y}$, that obey the equations
\begin{equation}\label{eq:connectors}
	\begin{cases}
	\nablas\uv_1=\uv_2\otimes\cv+\normal\otimes\dv_1,\\
		\nablas\uv_2=-\uv_1\otimes\cv+\normal\otimes\dv_2,\\
		\nablas\normal=-\uv_1\otimes\dv_1-\uv_2\otimes\dv_2,
	\end{cases}
\qquad
	\begin{cases}
		\nablast\vv_1=\vv_2\otimes\cv^\ast+\normal_{\y}\otimes\dv_1^\ast,\\
		\nablast\vv_2=-\vv_1\otimes\cv^\ast+\normal_{\y}\otimes\dv_2^\ast,\\
		\nablast\normal_{\y}=-\vv_1\otimes\dv_1^\ast-\vv_2\otimes\dv_2^\ast,
	\end{cases}
\end{equation}	
where $\nablast$ denotes the surface gradient on $\surface_{\y}$.
As in \eqref{eq:connectors_identity}, the curvature connectors must obey the identities
\begin{equation}
	\label{eq:connector_identities}
	\dv_1\cdot\uv_2=\dv_2\cdot\uv_1\quad\text{and}\quad\dv_1^\ast\cdot\vv_2=\dv_2^\ast\cdot\vv_1.
\end{equation}

Equations \eqref{eq:connectors} are instrumental to computing the tensor $\G$ starting from the following representation of $\R$,
\begin{equation}
	\label{eq:connectors_R}
	\R=\vv_1\otimes\uv_1+\vv_2\otimes\uv_2+\normal_{\y}\otimes\normal.
\end{equation}
To achieve this goal, we also need to combine \eqref{eq:connectors} with the equations
\begin{equation}
	\label{eq:connectors_chain_rule}
	\nablas\vv_i=(\nablast\vv_i)\nablas\y,\quad\nablas\normal_{\y}=(\nablast\normal_{\y})\nablas\y,
\end{equation}
which are mere applications of the chain rule. Moreover, we shall use the following representations for the connectors,
\begin{equation}
	\label{eq:connectors_components}
	\begin{cases}
		\cv=c_1\uv_1+c_2\uv_2,\\
		\dv_1=d_{11}\uv_1+d_{12}\uv_2,\\
		\dv_2=d_{12}\uv_1+d_{22}\uv_2,
	\end{cases}
\quad
\begin{cases}
	\cv^\ast=c_1^\ast\vv_1+c_2^\ast\vv_2,\\
	\dv_1^\ast=d_{11}^\ast\vv_1+d_{12}^\ast\vv_2,\\
	\dv_2^\ast=d_{12}^\ast\vv_1+d_{22}^\ast\vv_2,
\end{cases}
\end{equation}
where \eqref{eq:connector_identities} has also been used. By employing repeatedly \eqref{eq:connectors}, \eqref{eq:connector_identities}, \eqref{eq:connectors_chain_rule}, and \eqref{eq:connectors_components}, recalling that $\vv_i=\R\uv_i$, we arrive at the following form of \eqref{eq:G_representation_alternative},
\begin{equation}
	\label{eq:nablas_R}
	\G=\W_1\otimes\uv_1+\W_2\otimes\uv_2,
\end{equation}	
where $\W_i$ are skew-symmetric tensors defined by
\begin{equation}\label{eq:connectors_W_i}
\begin{aligned}
	\W_i&:=(\lambda_ic_i^\ast-c_i)\W(\normal)-(\lambda_id_{1i}^\ast-d_{1i})\W(\uv_2)+(\lambda_id_{i2}-d_{i2})\W(\uv_1)\\
	&\ =(\V\cv^\ast-\R\cv)\cdot\vv_i\W(\normal)-(\V\dv_1^\ast-\R\dv_1)\cdot\vv_i\W(\uv_2)+(\V\dv_2^\ast-\R\dv_2)\cdot\vv_i\W(\uv_1),
\end{aligned}
\end{equation}
and $\W(\uv_i)$ is the skew-symmetric tensor associated with $\uv_i$.

It follows from \eqref{eq:nablas_R} and \eqref{eq:connectors} that 
\begin{subequations}\label{eq:connectors_pre_averaged_contents}
\begin{align}
	\W(\normal)\circ\G&=(\W(\normal)\cdot\W_1)\uv_1+(\W(\normal)\cdot\W_2)\uv_2,\label{eq:connectors_pre_averaged_contents_d}\\
	|\G|^2&=\W_1\cdot\W_1+\W_2\cdot\W_2,\label{eq:connectors_pre_averaged_contents_b}\\
	\G\circ\nablas\normal&=-(\W_1\dv_1+\W_2\dv_2).\label{eq:connectors_pre_averaged_contents_c}
\end{align}	
\end{subequations}
Thus, by \eqref{eq:connectors_W_i}, we easily arrive at
\begin{subequations}\label{eq:connectors_averaged_contents}
\begin{align}
	|\W(\normal)\circ\G|^2&=4|\V\cv^\ast-\R\cv|^2,\label{eq:connectors_averaged_contents_d}\\
	|\nablas\G|^2-\frac12|\W(\normal)\circ\G|^2&=2\{|\V\dv_1^\ast-\R\dv_1|^2+|\V\dv_2^\ast-\R\dv_2|^2\},\label{eq:connectors_averaged_contents_b}\\
	\normal\cdot\G\circ\nablas\normal&=d_1^2+d_2^2-\V\dv_1^\ast\cdot\R\dv_1-\V\dv_2^\ast\cdot\R\dv_2.\label{eq:connectors_averaged_contents_c}
\end{align}
\end{subequations}

Equation \eqref{eq:connectors_averaged_contents_b} can also be given a possibly more telling form involving the  a mismatch between stretched and rotated curvature tensors. Since both curvature tensors $\nablas\normal$ and $\nablast\normal_{\y}$ are symmetric, they can also be written as
\begin{equation}
	\label{eq:connectors_curvature_tensors}
	\nablas\normal=-\dv_1\otimes\uv_1-\dv_2\otimes\uv_2,\quad\nablast\normal_{\y}=-\dv_1^\ast\otimes\vv_1-\dv_2^\ast\otimes\vv_2,
\end{equation}
and so 
\begin{equation}
	\label{eq:connectors_V_d-R_d}
	\V\dv_i^\ast-\R\dv_i=[\R\curvature\R\trans-\V(\nablast\normal_{\y})]\vv_i,
\end{equation}
which, once combined with \eqref{eq:connectors_averaged_contents_b}, delivers
\begin{equation}
	\label{eq:connectors_averaged_bending}
	|\G|^2-\frac12|\W(\normal)\circ\G|^2=2|\V(\nablast\normal_{\y})-\R\curvature\R\trans|^2.
\end{equation}	
Similarly, we arrive at
\begin{equation}
	\label{eq:connectors_averaged_extra}
	\normal\cdot\G\circ\nablas\normal=|\nablas\normal|^2-\V(\nablast\normal_{\y})\cdot\R\curvature\R\trans.
\end{equation}

By \eqref{eq:connectors_averaged_contents_d}, \eqref{eq:connectors_averaged_bending}, and \eqref{eq:connectors_averaged_extra} we can finally write the pure measures of drilling and bending envisioned  in our theory as
\begin{equation}
	\label{eq:pure_measures}
	\wdr=4|\V\cv^\ast-\R\cv|^2\quad\text{and}\quad\wb=4(|\V(\nablast\normal_{\y})|^2-|\nablas\normal|^2)^2.
\end{equation}
They vanish whenever, correspondingly,
\begin{equation}
	\label{eq:connectors_vanishing_energies}
	\V\cv^\ast=\R\cv\quad\text{or}\quad|\V(\nablast\normal_{\y})|=|\nablas\normal|.
\end{equation}
Informally, we may say that the former condition means that the stretched present spin connector equals the \emph{rotated} reference spin connector, while the latter means that the stretched present curvature tensor has the same strength as the  reference curvature tensor. It is worth noting that, by \eqref{eq:connectors_averaged_bending}, the bending measure  $\widetilde{w}_\mathrm{b}$ in \eqref{eq:w_b_plates} vanishes under a more restrictive requirement,
\begin{equation}
	\label{eq:restrictive_vanishing_requirement}
	\V(\nablast\normal_{\y})=\R\curvature\R\trans,
\end{equation} 
which says that the stretched present curvature tensor equals the rotated reference curvature tensor.\footnote{Were the rotation $\R$ replaced in \eqref{eq:restrictive_vanishing_requirement} by the full deformation gradient $\nablas\y$, \eqref{eq:restrictive_vanishing_requirement} would require the stretched present curvature tensor to coincide with the \emph{pushforward} of the reference curvature tensor, in complete analogy with the classical theory of Koiter~\cite{koiter:nonlinear}.}

It is  a simple matter to show that if $\y$ is a uniform rotation is space, $\y(\x)=\Q\x$ with $\Q\in\orth$ independent of $\x$, then the equalities in both \eqref{eq:connectors_vanishing_energies} and \eqref{eq:restrictive_vanishing_requirement} are identically satisfied, as $\V=\Proj(\normal_{\y})$, $\cv^\ast=\Q\cv$, and $\nablast\normal_{\y}=\Q\curvature\Q\trans$.

\section{Conclusion}\label{sec:conclusion}
We have proposed a direct hyperelastic theory for soft shells, based upon a separation criterion that requires the strain-energy density (per unit area in the reference configuration) to be the sum of \emph{three} independent contributions, each penalizing a single deformation mode. Alongside the traditional stretching deformation mode, we have identified and described two non-stretching modes, that is, \emph{drilling} and \emph{bending}. They originate from the distinct, uniquely identified rotations embedded in the deformation gradient that have axes along the local normal to the reference surface and on the tangent plane, respectively. Invariant averages were extracted out of these frame-dependent kinematic contents by a method applicable when the material is transversely isotropic about the normal. These averages turned out to depend on the second deformation gradient via an invariant third-order tensor $\G$ constructed from the surface gradient of the polar rotation $\R$ (extracted from the polar decomposition of the first deformation gradient).

We identified local pure drilling and pure bending deformations upon which pure bending and pure drilling measures derived from the corresponding invariant averages were required to vanish (in that order) to enforce our separation criterion. We found a pure measure of drilling quadratic in $\G$, but a pure measure of bending must in general be at least \emph{quartic}, although it could be taken quadratic in special cases (notably, for plates and cylindrical shells).

Although our pure measures emerged from a formal reasoning, originally divorced from geometry, they were also shown to be related to mismatches between geometric quantities evaluated in the deformed and undeformed configurations of the shell, including the curvature tensor. 

The stretching tensor $\V$, which is the primary player in pure stretching measures, also features in our pure measures of drilling and bending: this indeed turns out to be necessarily the case for a pure measure to be such in our theory, as was already in the elastic theory of Antman for curved rods in \cite{antman:general}.

Of the local pure drilling and bending deformations that were instrumental to the definition of the ground state for the strain energy, the former are also known to exist in the large; they are the \emph{bending-neutral} deformations studied in \cite{sonnet:bending-neutral}. On the other hand, pure bending deformations are not known to exist in the large (apart from the special case of cylindrical shells). 

We also encountered a peculiar case of \emph{soft elasticity} predicted by our theory, where all pure measures vanish on a family of deformations which produced a class of isometric minimal surfaces (see Appendix~\ref{sec:soft_elasticity}). One may wonder whether this is indeed the sign of a more general property and ask: are all isometric bending-neutral deformations of a minimal surface cases of soft elasticity for our theory?

We have called \emph{soft} the shells described by our theory to highlight the role played here by the drilling deformation mode, which is expected to remain silent for \emph{hard} structural shells. The softest possible shell is perhaps a soap bubble, which at equilibrium would acquire the shape of a minimal surface. Were all minimal surfaces in the ground state of our theory, a soap bubble would freely glide from an equilibrium shape to another, if only allowed by the enforced boundary conditions. This issue has not been addressed here.

Material symmetry was not addressed either. Apart from the requirement of transverse isotropy, which we believe is implicit in our averaging process, systematic considerations of material symmetry, in the same spirit as those expounded for other direct second-grade theories in \cite{steigmann:equilibrium,murdoch:symmetry,davini:material} were not entertained. Presumably, different material symmetries would affect our averaging process, making $\ave{d^2}$ and $\ave{b^2}$ more openly akin to order parameters.

\appendix

\section{Surface Calculus: Definitions and Calculations}\label{sec:surface_calculus}
Here, we collect a few definitions and properties that apply to calculus on a smooth surface $\surface$.

Let $\varphi:\surface\to\mathbb{R}$ be a scalar field; we say that $\varphi$ is differentiable at a point identified on $\surface$ by the position vector $\x$, if there is a vector $\nablas\varphi$ on the tangent plane $\tplane_{\x}$ to $\surface$ at $\x$ such that for every curve $\curve:t\mapsto\curve(t)\in\surface$ designating a trajectory on $\surface$ for which $\curve(t_0)=\x$,
\begin{equation}\label{eq:surface_gradient_scalar}
	\left.\frac{\dd}{\dd t}\varphi(\curve(t))\right|_{t=t_0}=\nablas\varphi\cdot\tangent,
\end{equation}
where $\tangent$ is the unit tangent vector to $\curve$ at $\x$. We call $\nablas\varphi$ the surface gradient of $\varphi$. Similarly, a vector field $\h:\surface\to\transl$ is differentiable at $\x\in\surface$, if 
\begin{equation}\label{eq:surface_gradient_vector}
	\left.\frac{\dd}{\dd t}\h(\curve(t))\right|_{t=t_0}=(\nablas\h)\tangent,
\end{equation}
where the surface gradient $\nablas\h$ is a second-rank tensor mapping $\tplane_{\x}$ into $\transl$. We can also differentiate a second-rank tensor field $\F:\surface\to\lin$, where $\lin$ is the \emph{linear group} on the three-dimensional translation space $\transl$. The surface gradient $\nablas\F$, defined by
\begin{equation}
	\label{eq:surface_gradient_tensor}
	\left.\frac{\dd}{\dd t}\F(\curve(t))\right|_{t=t_0}=(\nablas\F)\tangent,
\end{equation}
is a third-rank tensor mapping $\tplane_{\x}$ into $\lin$. 

Letting $\normal$ denote a unit normal field orienting $\surface$, if $\nablas\h$ is \emph{dyadic}, that is, $\nablas\h=\av_1\otimes\av_2$, then $\av_2\cdot\normal=0$ and $(\nablas\h)\tangent=(\av_2\cdot\tangent)\av_1$; similarly, if $\nablas\F$ is \emph{triadic}, that is, $\nablas\F=\av_1\otimes\av_2\otimes\av_3$, then $\av_3\cdot\normal=0$ and $(\nablas\F)\tangent=(\av_3\cdot\tangent)\av_1\otimes\av_2$.

The surface gradients $\nablas\h$ and $\nablas\F$ that we have encountered in the main text can always be seen as linear combinations of dyadic and triadic tensors.
In particular, given a Cartesian frame $\framee$, we can write
\begin{equation}
	\label{eq:component_representation}
	\F=F_{ij}\e_i\otimes\e_j\quad\text{and}\quad\nablas\F=F_{ij;k}\e_i\otimes\e_j\otimes\e_k,
\end{equation}
where a semicolon denotes differentiation with respect to tangential components.\footnote{The convention of summing over repeated indices is adopted whenever components are used.}
Similarly, we can decompose the third-rank tensor $\G$ defined by \eqref{eq:G_definition} as a linear combination,
\begin{equation}
	\label{eq:G_linear_combination}
	\G=G_{ijk}\e_i\otimes\e_j\otimes\e_k,
\end{equation}
where the scalar components $G_{ijk}$ obey the conditions
\begin{equation}
	\label{eq:G_component_conditions}
	G_{ijk}=-G_{jik}\quad\text{and}\quad G_{ijk}\nu_k=0,
\end{equation}
the latter following from \eqref{eq:G_definition} when denoting by $\nu_k$ the components of $\normal$ in the given frame. Similarly, the skew-symmetric tensor $\W_{\uv}:=\G\uv$, where $\uv$ is a unit vector such that
\begin{equation}
	\label{eq:orthogonality}
	\uv\cdot\normal=0,
\end{equation}
has the following representation,
\begin{equation}
	\label{eq:W_u_components}
	\W_{\uv}=W_{ij}\e_i\otimes\e_j\quad\text{with}\quad W_{ij}:=G_{ijk}u_k,
\end{equation}
where $u_k$ are the components of $\uv$.

Letting any second-rank tensor $\A$ be represented as
\begin{equation}
	\label{eq:A_components}
	\A=A_{ij}\e_i\otimes\e_j,
\end{equation}
we can write
\begin{equation}
	\label{eq:W_A}
	\W_{\uv}\cdot\A=W_{ij}A_{ij}=A_{ij}G_{ijk}u_k,
\end{equation}
so that 
\begin{equation}
	\label{eq:W_A_averaged}
	\ave{(\W_{\uv}\cdot\A)^2}_{\uv\in\disc}=A_{ij}G_{ijk}A_{lm}G_{lmh}\ave{u_ku_h}_{\uv\in\disc}=\frac12A_{ij}G_{ijk}A_{lm}G_{lmk},
\end{equation}
where we have used \eqref{eq:orthogonality} and the following form of the identity \eqref{eq:average_identity},
\begin{equation}
	\label{eq:average_identity_components}
	\ave{u_ku_h}_{\uv\in\disc}=\frac12(\delta_{kh}-\nu_k\nu_h),
\end{equation}
where $\delta_{kh}$ is Kronecker's symbol. Equation \eqref{eq:averaged_drilling} in the main text is then an immediate consequence of \eqref{eq:W_A_averaged} and definition \eqref{eq:generic_triadic}.

To prove \eqref{eq:averaged_bending}, we only need to remark that 
\begin{equation}
	\label{eq:W_W_components}
	\W_{\uv}\cdot\W_{\uv}=G_{ijk}G_{ijh}u_ku_h\quad\text{and}\quad |\G|^2=G_{ijk}G_{ijk},
\end{equation}
and apply again \eqref{eq:average_identity_components}. 

\section{Bending-Neutrality}\label{sec:bending_neutral}
Here, we prove that applying a pure drilling deformation to an already deformed surface leaves its bending content unchanged.

Assume that a surface $\surface$ has been deformed by $\y^\ast$ into the surface $\surface^\ast$, which is further deformed by a pure drilling deformation $\y'$ into the surface $\surface'=\y'(\surface^\ast)$. We denote by $\R^\ast$ the polar rotation associated with $\nablas\y^\ast$ as in \eqref{eq:polar_decomposition}, and by $\R'$ the polar rotation associated with $\nablast\y'$, where $\nablast$ is the surface gradient on $\surface^\ast$. Moreover, $\normal^\ast$ and $\normal'$ are the unit normal fields on $\surface^\ast$ and $\surface'$, respectively.

Let $\y:=\y^\ast\circ\y'$ be the composed deformation that changes $\surface$ into $\surface'$ in a single stroke. Since in the present frame $\y'$ preserves the normal to $\surface^\ast$, 
\begin{equation}
	\label{eq:bending_netral_normals}
	\normal'=\R'\normal^\ast=\normal^\ast=\R^\ast\normal,
\end{equation}
where $\normal$ is the unit normal field to $\surface$. Letting $\R$ denote the polar rotation extracted from $\nablas\y$, \eqref{eq:bending_netral_normals} can also be rewritten as 
\begin{equation}
	\label{eq:bending_netral_rotations}
	\R\normal=\R^\ast\normal,
\end{equation}
from which, decomposing both $\R$ and $\R^\ast$ in their bending and drilling components as in \eqref{eg:rotation_decomposition}, here denoted $\Rb$, $\Rb^\ast$ and $\Rd$, $\Rd^\ast$, respectively, we obtain that
\begin{equation}
	\label{eq:bending_neutral_decomposition}
	\Rb\Rd\normal=\Rb\normal=\Rb^\ast\Rd^\ast\normal=\Rb^\ast\normal.
\end{equation}
Calling $\bv$ and $\bv^\ast$ the bending contents of $\y$ and $\y^\ast$, respectively, and using \eqref{eq:rodrigues_formula} in \eqref{eq:bending_neutral_decomposition}, since both $\bv$ and $\bv^\ast$ are vectors tangent to $\surface$, we arrive at 
\begin{equation}
	\label{eq:bending_neutral_conclusion}
	\frac{1}{1+b^2}((1-b^2)\normal+2\bv\times\normal)=\frac{1}{1+{b^\ast}^2}((1-{b^\ast}^2)\normal+2\bv^\ast\times\normal).
\end{equation}
Projecting both sides of this equation along $\normal$ and on the plane tangent to $\surface$, delivers 
\begin{equation}
	\label{eq:bending_neutral_desired}
	\bv=\bv^\ast,
\end{equation}
which is the desired result.

\section{Contents' Frame-Dependence}\label{sec:contents_not_measures}
Since in a change of frame, the polar rotation $\R$ is transformed into $\R^\ast=\Q\R$, neither of the vectors $\dv$ or $\bv$ is frame-invariant or frame-indifferent. Here, we explore by example how the vectors' lengths $a$, $d$, and $b$ transform under a change of frame. In particular, we shall show how the angles $\alpha$, $\alpha_\mathrm{d}$, and $\alpha_\mathrm{b}$ appear in a family of frames parameterized by a single rotation angle. 

\begin{figure}[]
	\centering
	\begin{subfigure}[c]{0.55\textwidth}
		\centering
		\includegraphics[width=\textwidth]{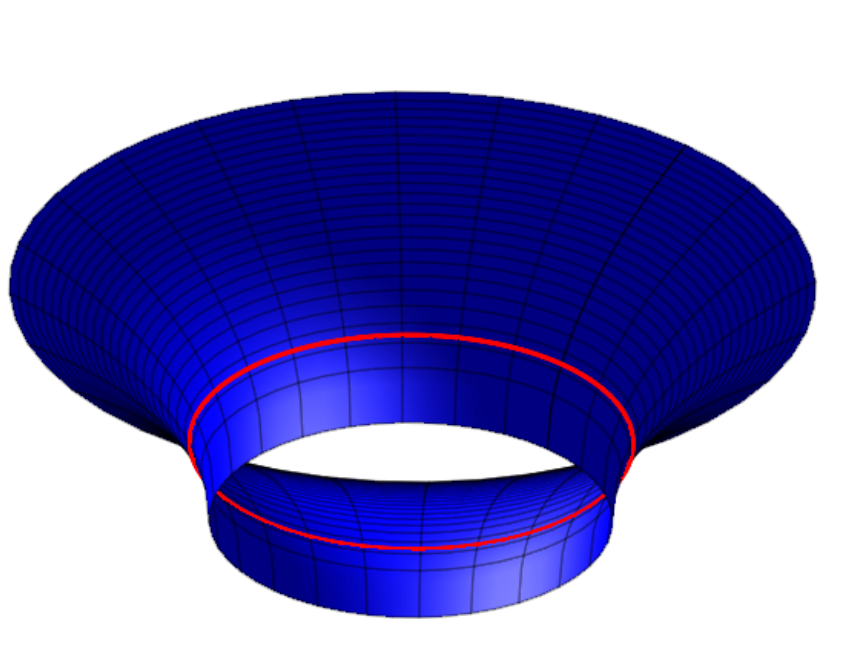}
		\caption{The half-catenoid described by \eqref{eq:catenoid} for $0\leqq\vartheta\leqq2\pi$ and $a\leqq r\leqq2a$. The neck has radius $a$, while the red circle has radius $r\doteq1.09a$.}
		\label{fig:catenoid}
	\end{subfigure}
	\hfill
	\begin{subfigure}[c]{0.35\textwidth}
		\centering
		\includegraphics[width=\textwidth]{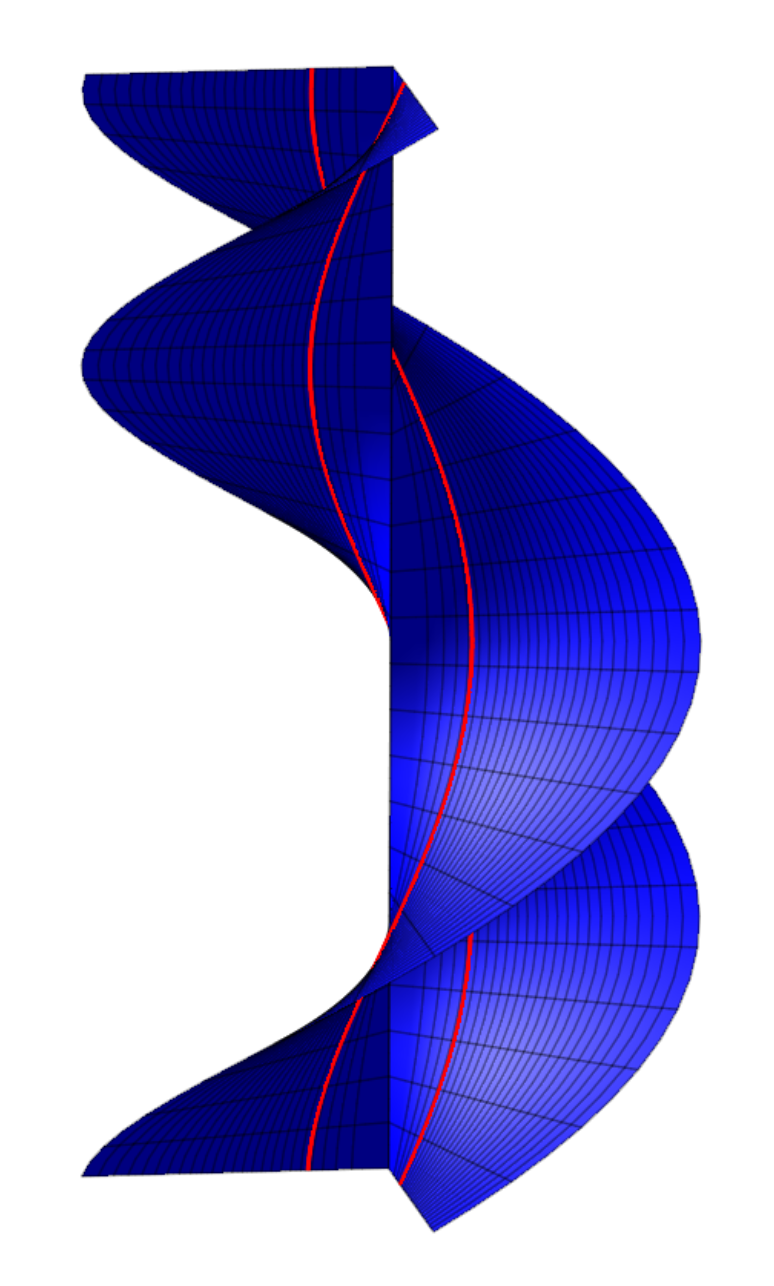}
		\caption{Half-helicoids obtained by deforming the half-catenoid in Fig.~\ref{fig:catenoid} through $\y$ in \eqref{eq:catenoid_deformation} with $\vartheta_0=\pi/4$ and $\vartheta_0=3\pi/4$. The red helices are images of the red circle in Fig.~\ref{fig:catenoid}. The two surfaces differ only by a rotation of angle $\pi/2$ about the helicoids' common axis.}
		\label{fig:helicoids}
	\end{subfigure}
	\hfill
	\\
	\begin{subfigure}[c]{0.45\textwidth}
		\centering
		\includegraphics[width=\textwidth]{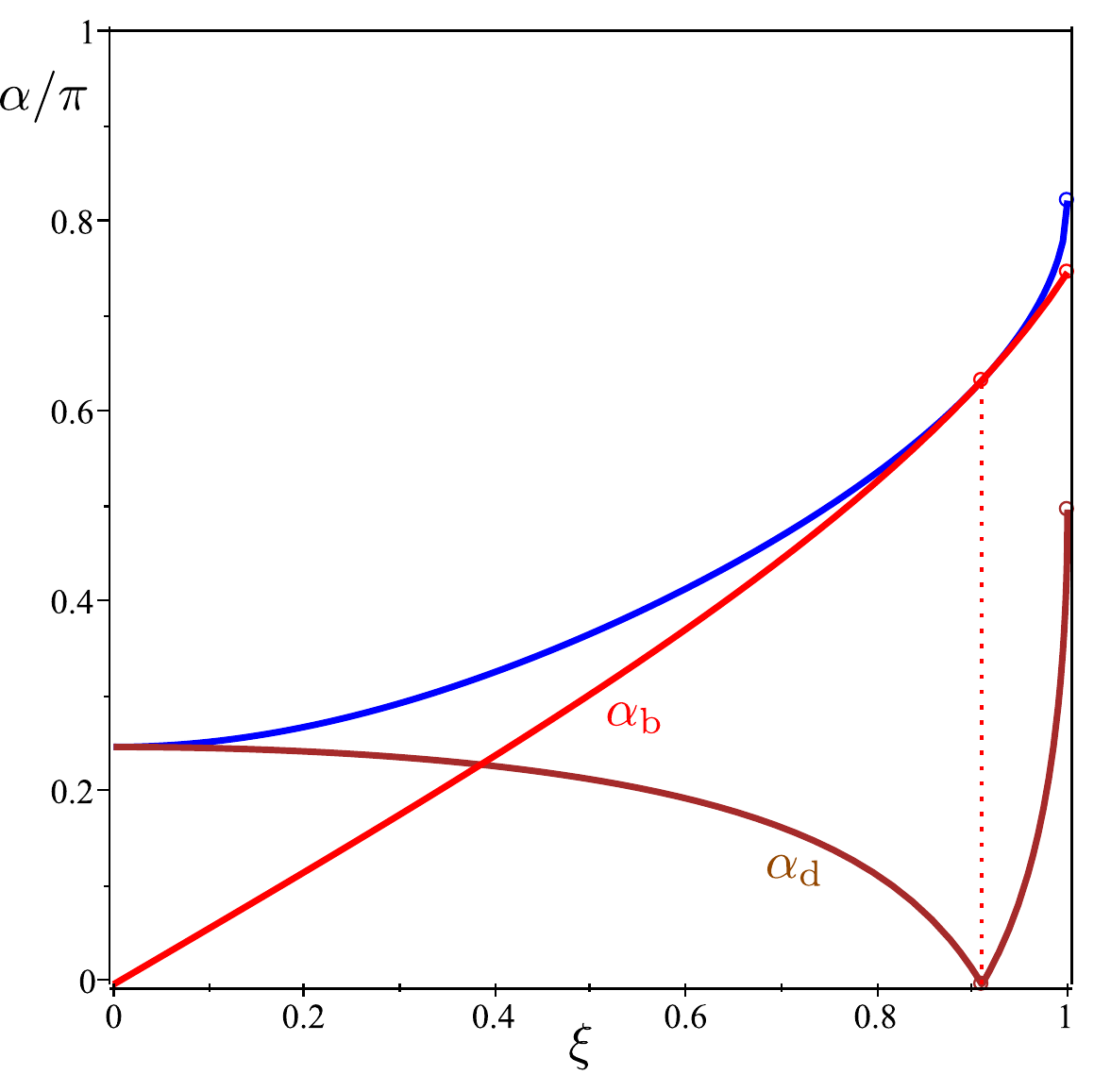}
		\caption{Plots of $\alpha$, $\alpha_\mathrm{d}$, and $\alpha_\mathrm{b}$ extracted from \eqref{eq:catenoid_a_d_b} as functions of $\xi:=a/r$ for $\vartheta_0=\pi/4$. The drilling angle $\alpha_\mathrm{d}$ vanishes for $\xi\doteq0.91$, corresponding to the red circle in Fig.~\ref{fig:catenoid}.}
		\label{fig:angles_pi_over_4}
	\end{subfigure}
	\hfill
	\begin{subfigure}[c]{0.45\textwidth}
		\centering
		\includegraphics[width=\textwidth]{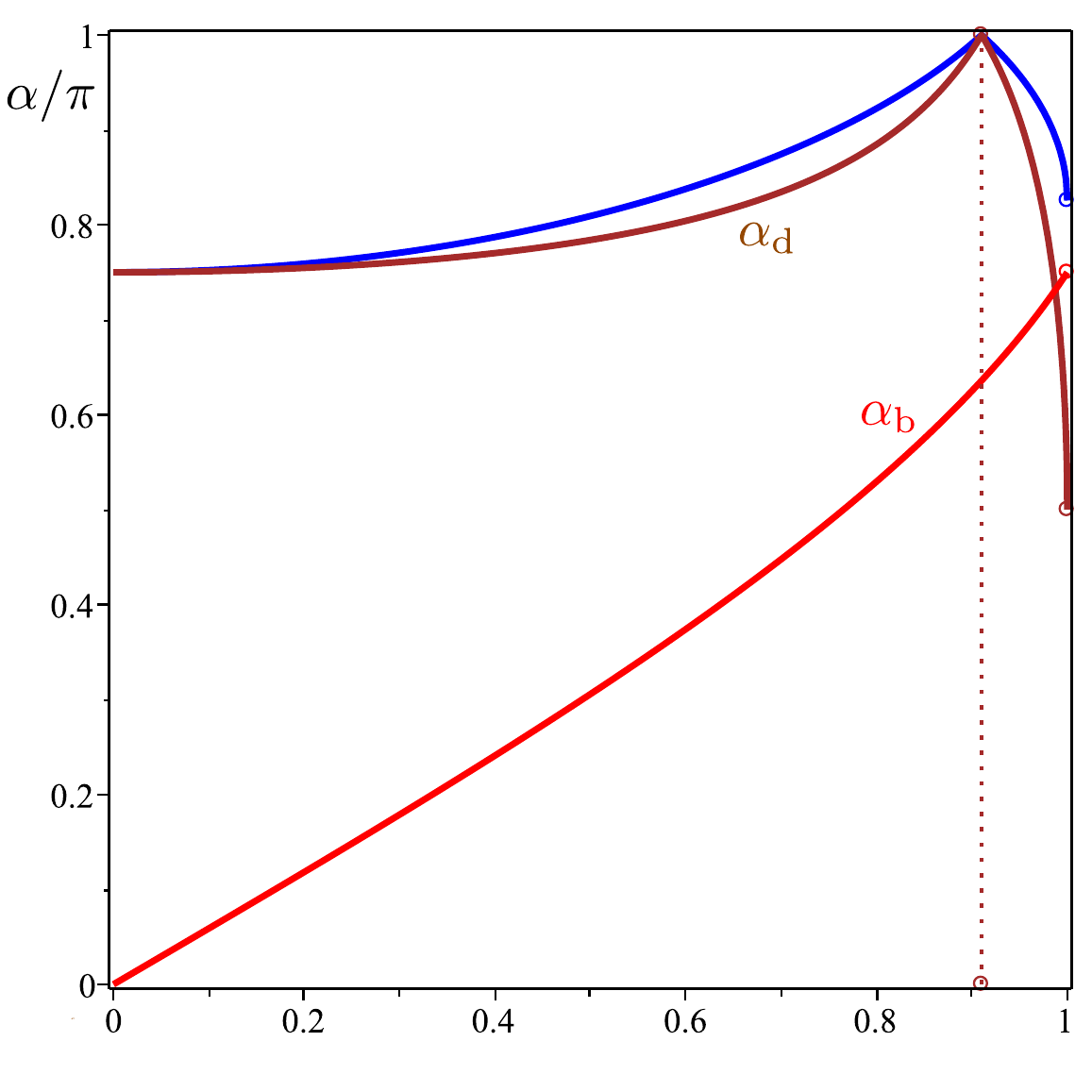}
		\caption{Plots of $\alpha$, $\alpha_\mathrm{d}$, and $\alpha_\mathrm{b}$ extracted from \eqref{eq:catenoid_a_d_b} as functions of $\xi:=a/r$ for $\vartheta_0=3\pi/4$. The drilling angle $\alpha_\mathrm{d}$ equals $\pi$ for $\xi\doteq0.91$, corresponding to the red circle in Fig.~\ref{fig:catenoid}.}
		\label{fig:angles_3_pi_over_4}
	\end{subfigure}
	\hfill
	\caption{An overall rotation about the axis $\e_3$ affects dramatically the drilling and bending angles. Only for $\vartheta_0=-\pi/2$ do these reduce to $\alpha_\mathrm{d}=\pi/2$ and $\alpha_\mathrm{b}=0$, indicating that $\y$ in \eqref{eq:catenoid_deformation} is indeed a bending-neutral deformation corresponding to the classical Bonnet transformation of a catenoid into a helicoid.}
	\label{fig:overall_rotation}
\end{figure}

Let $\surface$ be the half-catenoid depicted in Fig.~\ref{fig:catenoid}; in a Cartesian frame $\framee$, it can be represented  as (see also \cite{chen:isometries}),
\begin{equation}
	\label{eq:catenoid}
	\x(r,\vartheta)=r\cos\vartheta\e_1+r\sin\vartheta\e_2+a\arccosh\left(\frac{r}{a}\right)\e_3,
\end{equation}
where $0\leqq\vartheta\leqq2\pi$, $r\geqq a$, and $a>0$ is the radius of the \emph{neck}.
We scale all lengths to $a$ and introduce the dimensionless variable $\xi:=a/r$, which ranges in the interval $[0,1]$. The surface $\surface_{\y}$ is the half-helicoid delivered by the deformation $\y:\x\mapsto\y(\x)$ represented in the same parametric form as $\x$ by
\begin{equation}
	\label{eq:catenoid_deformation}
	\y(r,\vartheta)=a\left\{\frac{1}{\xi}\sqrt{1-\xi^2}\cos(\vartheta+\vartheta_0)\e_1+\frac{1}{\xi}\sqrt{1-\xi^2}\sin(\vartheta+\vartheta_0)\e_2+\vartheta\e_3\right\},
\end{equation}
where $\vartheta_0$ is a parameter chosen in the interval $[-\pi,\pi]$. All surfaces $\surface_{\y}$ represented by \eqref{eq:catenoid_deformation} differ from one another by the action of a rigid (counter-clockwise) rotation $\Q$ by angle $\vartheta_0$ about the axis $\e_3$.

For a given $\vartheta_0$, we now compute $\nablas\y$. This task is better accomplished in the mobile frame $(\e_r,\e_\vartheta,\normal)$ on $\surface$, where $\normal:=\e_r\times\e_\vartheta$ and the unit vectors $(\e_r,\e_\vartheta)$ are associated with the variables $(r,\vartheta)$ so that, for any curve $t\mapsto\x(t)$ on $\surface$, we can decompose $\dot{\x}$ as
\begin{equation}
	\label{eq:x_dot}
	\dot{\x}=\frac{\dot{r}}{\sqrt{1-\xi^2}}\e_r+a\frac{\dot{\vartheta}}{\xi}\e_\vartheta.
\end{equation}
Simple calculations show that
\begin{equation}\label{eq:movable_frame}
	\begin{cases}
	\begin{aligned}
		\e_r&=\sqrt{1-\xi^2}(\cos\vartheta\e_1+\sin\vartheta\e_2)+\xi\e_3,\\
		\e_\vartheta&=-\sin\vartheta\e_1+\cos\vartheta\e_2\\
		\normal&=-\xi(\cos\vartheta\e_1+\sin\vartheta\e_2)+\sqrt{1-\xi^2}\e_3.
	\end{aligned}
\end{cases}
\end{equation}
By use of \eqref{eq:catenoid_deformation} and \eqref{eq:movable_frame}, we easily see that the curve $t\mapsto\x(t)$ on $\surface$ is transformed into a curve $t\mapsto\y(t)$ on $\surface_{\y}$, for which
\begin{align}
	\label{eq_y_dot}
	\dot{\y}=a\bigg\{&\left(\left(\xi-\frac{1-\xi^2}{\xi}\sin\vt_0\right)\dot{\vt}+\cos\vt_0\frac{\dot{r}}{a}\right)\e_r+\left(\frac{\sqrt{1-\xi^2}}{\xi}\cos\vt_0\dot{\vt}+\frac{1}{\sqrt{1-\xi^2}}\sin\vt_0\frac{\dot{r}}{a}\right)\e_\vartheta\nonumber\\
	&+\left(\sqrt{1-\xi^2}(1+\sin\vt_0)\dot{\vartheta}-\frac{\xi}{\sqrt{1-\xi^2}}\cos\vt_0\frac{\dot{r}}{a}\right)\normal\bigg\}.
\end{align}
By requiring that $\dot{\y}=(\nablas\y)\dot{\x}$ for all possible curves $\x(t)$, we obtain that
\begin{equation}
	\begin{aligned}
		\label{eq:catenoid_deformation_gradient}
		\nablas\y&=\sqrt{1-\xi^2}\cos\vartheta_0\e_r\otimes\e_r+(\xi^2+(\xi^2-1)\sin\vartheta_0)\e_r\otimes\e_\vartheta+\sin\vartheta_0\e_\vartheta\otimes\e_r\\&+\sqrt{1-\xi^2}\cos\vartheta_0\e_\vartheta\otimes\e_\vartheta-\xi\cos\vartheta_0\normal\otimes\e_r+\xi\sqrt{1-\xi^2}(1+\sin\vartheta_0)\normal\otimes\e_\vartheta,
	\end{aligned}
\end{equation}
whence it follows that $\U=\proj$, so that the deformation $\y$ in \eqref{eq:catenoid_deformation} is an isometry for all $\vartheta_0$, and 
\begin{equation}
	\label{eq:catenoid_cofactor}
	\co((\nablas\y)\proj)=\xi\sqrt{1-\xi^2}(1+\sin\vartheta_0)\e_r\otimes\normal-\xi\cos\vartheta_0\e_\vartheta\otimes\normal+(1-\xi^2(1+\sin\vartheta_0))\normal\otimes\normal.
\end{equation}

In a similar way, it follows from \eqref{eq:x_dot} and \eqref{eq:movable_frame} that 
\begin{equation}\label{eq:catenoid_movable_frame}
	\begin{cases}
		\begin{aligned}
	\nablas\e_r&=\frac{1}{a}\Big(\xi\sqrt{1-\xi^2}\e_\vartheta\otimes\e_\vartheta-\xi^2\normal\otimes\e_r\Big),\\
	\nablas\e_\vt&=\frac{1}{a}\Big(-\xi\sqrt{1-\xi^2}\e_r\otimes\e_\vartheta+\xi^2\normal\otimes\e_\vt\Big),\\
	\nablas\normal&=\frac{1}{a}\Big(\xi^2\e_r\otimes\e_r-\xi^2\e_\vartheta\otimes\e_\vartheta\Big),
	\end{aligned}
	\end{cases}
\end{equation}
so that the connectors of the movable frame $(\e_r,\e_\vt,\normal)$ of a catenoid are
\begin{equation}
	\label{eq:catenoid_connectors}
	\cv=\frac{\xi}{a}\sqrt{1-\xi^2}\e_\vt,\quad\dv_1=-\frac{\xi^2}{a}\e_r,\quad\dv_2=\frac{\xi^2}{a}\e_\vt.
\end{equation}

Inserting both \eqref{eq:catenoid_deformation_gradient} and \eqref{eq:catenoid_cofactor} in \eqref{eq:rotation_extraction}, 
by use of \eqref{eq:a^2} and \eqref{eq:d_and_b_scalars_explicit}, we arrive at
\begin{subequations}	\label{eq:catenoid_a_d_b}
	\begin{align}
		a^2&=\frac{2(1-\sqrt{1-\xi^2}\cos\vartheta_0)+\xi^2(1+\sin\vartheta_0)}{2(1+\sqrt{1-\xi^2}\cos\vartheta_0)-\xi^2(1+\sin\vartheta_0)},\label{eq:catenoid_a}\\
		d^2&=\left(\frac{2\sin\vartheta_0-\xi^2(1+\sin\vartheta_0)}{2(1+\sqrt{1-\xi^2}\cos\vartheta_0)-\xi^2(1+\sin\vartheta_0)} \right)^2,\label{eq:catenoid_d}\\
		b^2&=\frac{\xi^2(1+\sin\vartheta_0)}{2-\xi^2(1+\sin\vartheta_0)},\label{q:catenoid_b}
	\end{align}
\end{subequations}
which clearly depend on the change of frame embodied by the rotation $\Q$ by angle $\vt_0$ about $\e_3$.
As an illustration of these equations, we plot in Figs.~\ref{fig:angles_pi_over_4} and \ref{fig:angles_3_pi_over_4} the angles $\alpha$, $\alpha_\mathrm{d}$, and $\alpha_\mathrm{b}$ as functions of $\xi$ for both $\vartheta_0=\pi/4$ and $\vartheta_0=3\pi/4$. In particular, one sees how in a frame the drilling angle could vanish along a circle of $\surface$, precisely the same where it attains its maximum in a different frame, rotated relative to the other by angle $\pi/2$ about $\e_3$  (see Fig.~\ref{fig:helicoids}).

There is one special frame where $b^2=0$ and $d^2=a^2$, so that $\alpha_\mathrm{b}=0$ and $\alpha=\alpha_\mathrm{d}=\pi/2$: it is obtained for $\vt_0=-\pi/2$. This indicates that  $\y$ in \eqref{eq:catenoid_deformation} is indeed a bending-neutral deformation corresponding to the classical Bonnet transformation of a catenoid (see also \cite{sonnet:bending-neutral}).

Clearly, neither $d^2$ nor $b^2$ is an invariant measure of drilling or bending.

\section{Soft Elasticity}\label{sec:soft_elasticity}
Had we not prescribed a pure bending deformation to have the direction of bending locally constant, the definition of $\wdr$  in \eqref{eq:w_d} would not be acceptable as a measure of pure drilling, as it would not vanish on all pure bending deformations. Here we want to define an alternative measure of pure drilling $\widetilde{w}_\mathrm{d}$ that vanishes on all unconstrained pure bending deformations. To this end, we remark that, by \eqref{eq:G_b},
\begin{equation}
	\label{P_G_W}
	\proj(\G\circ\W(\normal))=(1-\cos\beta)\cv+\sin\beta\dv_2,
\end{equation}
so that, by \eqref{eq:W_G_b} and in tune with \eqref{eq:w_b}, a natural definition for $\widetilde{w}_\mathrm{d}$ would be
\begin{equation}
	\label{eq:w_d_tilde}
	\widetilde{w}_\mathrm{d}:=|\W(\normal)\circ\G|^2-4|\proj(\G\circ\W(\normal))|^2.
\end{equation}
However, it follows from the general representation for $\G$ in \eqref{eq:G_representation_alternative} that
\begin{equation}
	\label{eq:G_identity}
	\proj(\G\circ\W(\normal))=-\frac12\W(\normal)\circ\G,
\end{equation}
and so $\widetilde{w}_\mathrm{d}$ vanishes identically.

Thus, according to our theory, enlarging the class of pure bending deformations would entail that  no pure measure of drilling could exist. This would mean that no energy cost could ever be associated with a pure drilling mode, which would then be universally \emph{soft}. 
Since we are not prepared to accept such a general conclusion, we have adopted $\wdr$ in \eqref{eq:w_d} as a pure measure of drilling. However, even with this choice for $\wdr$,  special instances of \emph{soft elasticity} still remain possible: these are deformations that make $W$ in \eqref{eq:W} vanish identically.

\begin{figure}[]
	\centering\includegraphics[width=.33\linewidth]{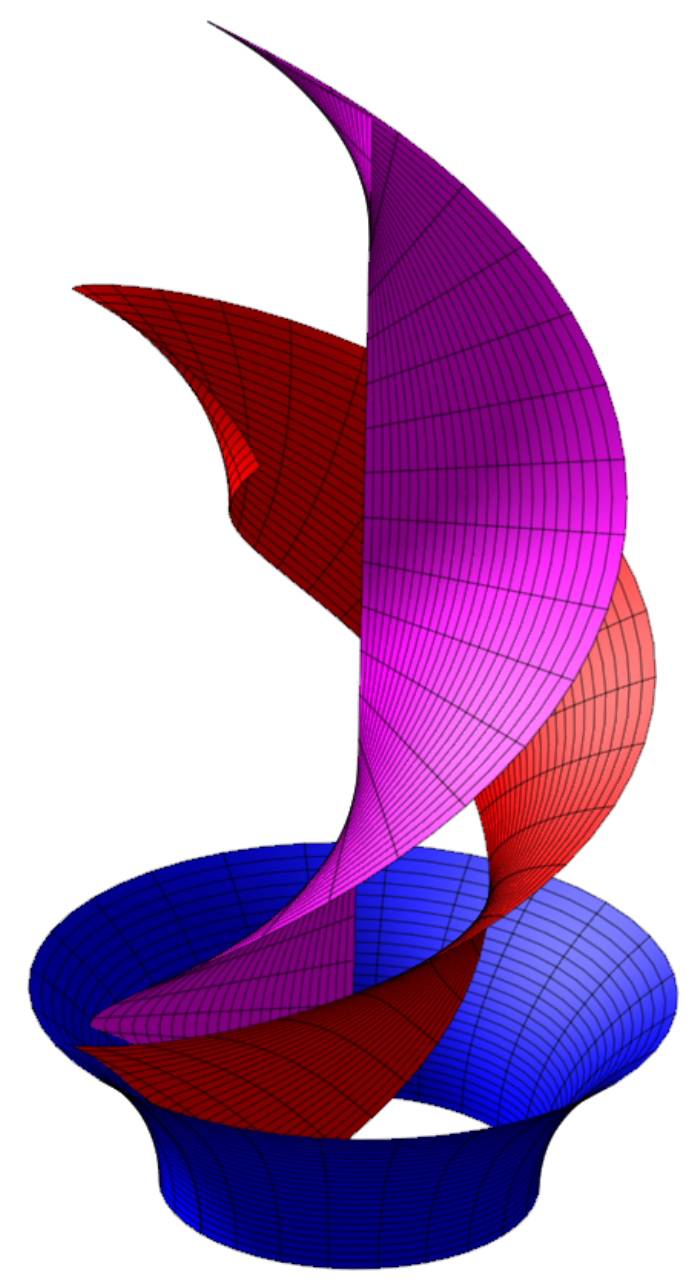}
	\caption{The surface $\y_\alpha(\surface)$ represented by \eqref{eq:y_alpha}, with $\xi$ ranging in $[\frac12,1]$ and $\vt$ in $[0,2\pi]$, for $\alpha=0$ (blue), $\alpha=\frac{\pi}{4}$ (red), and $\alpha=\frac{\pi}{2}$ (magenta). The half-catenoid corresponding to $\alpha=0$ is the same as in Fig.~\ref{fig:catenoid}.}
	\label{fig:scherk}
\end{figure}

Consider again as a reference surface $\surface$ the half-catenoid represented by \eqref{eq:catenoid} and subject it to the deformation described by
\begin{equation}
	\label{eq:y_alpha}
\begin{split}
\y_\alpha(r,\vt):=a\bigg\{&\cos\alpha\bigg(\frac{1}{\xi}\cos\vt\e_1+\frac{1}{\xi}\sin\vt\e_2+\arccosh\bigg(\frac{1}{\xi}\bigg)\e_3\bigg)\\
&+\sin\alpha\bigg(\frac{1}{\xi}\sqrt{1-\xi^2}\sin\vt\e_1-\frac{1}{\xi}\sqrt{1-\xi^2}\cos\vt\e_2+\vt\e_3\bigg)\bigg\},
\end{split}
\end{equation}
where $\xi=a/r\leqq1$ and $\alpha$ is a constant ranging in $[0,\frac{\pi}{2}]$. For $\alpha=0$, $\y_\alpha$ is the identity, and so $\y_\alpha(\surface)$ is the catenoid $\surface$ itself, whereas for $\alpha=\frac{\pi}{2}$, $\y_\alpha$ is the same as the deformation defined by \eqref{eq:catenoid_deformation} for $\vt_0=-\frac{\pi}{2}$, and so $\y_\alpha(\surface)$ is a half-helicoid. For $0<\alpha<\frac{\pi}{2}$, $\y_\alpha(\surface)$ is also a minimal surface, precisely, it is \emph{Scherk's second surface} (see, for example, p.\,148 of \cite{dierkes:minimal}). Figure~\ref{fig:scherk} illustrates the surface $\y_\alpha(\surface)$ for three values of $\alpha$.

By use of \eqref{eq:catenoid_deformation_gradient} with $\vt_0=-\frac{\pi}{2}$, it is a simple matter to show that
\begin{equation}
	\label{eq:soft_deformation_gradient}
	\nablas\y_\alpha=\cos\alpha\proj-\sin\alpha\W(\normal),
\end{equation}
whence it follows that $(\nablas\y_\alpha)\trans(\nablas\y_\alpha)=\proj$, so that $\y_\alpha$ is an isometry for all values of $\alpha$ and $\ws$ in \eqref{eq:w_s} vanishes. Then we extract from $\nablas\y_\alpha$ its polar rotation,
\begin{equation}
	\label{eq:soft_rotation}
	\R_\alpha=\cos\alpha\I+(1-\cos\alpha)\normal\otimes\normal-\sin\alpha\W(\normal),
\end{equation}
which, by \eqref{eq:euler_formula}, represents a \emph{clockwise} rotation by angle $\alpha$ about $\normal$. A simple calculation that uses \eqref{eq:catenoid_movable_frame} leads us to
\begin{equation}
	\label{eq:G_alpha}
	\begin{split}
		\G_\alpha:=\R_\alpha\trans\nablas\R_\alpha=
		\frac{\xi^2}{a}\bigg\{&-\bigg((1-\cos\alpha)\W(\e_\vt)+\sin\alpha\W(\e_r)\bigg)\otimes\e_r\\&+\bigg(\sin\alpha\W(\e_\vt)-(1-\cos\alpha)\W(\e_r) \bigg)\otimes\e_\vt\bigg\},
	\end{split}
\end{equation}
from which it follows that
\begin{equation}
	\label{eq:soft_measures}
	|\G_\alpha|^2=\frac{8\xi^4}{a^2}(1-\cos\alpha),\quad\W(\normal)\circ\G_\alpha=\zero,\quad\G_\alpha\circ\nablas\normal=\frac{2\xi^4}{a^2}(1-\cos\alpha)\normal,
\end{equation}
so that both $\wdr$ and $\wb$, as given by \eqref{eq:w_d} and \eqref{eq:w_b}, respectively, vanish. 

Thus we conclude that within the theory presented in this paper the special class of deformations $\y_\alpha$ in \eqref{eq:y_alpha} represents a case of \emph{soft elasticity} for the energy density $W$ in \eqref{eq:W}.

\section{Cylindrical Shells}\label{sec:cylindrical_shells}
\begin{figure}[]
	\centering\includegraphics[width=.75\linewidth]{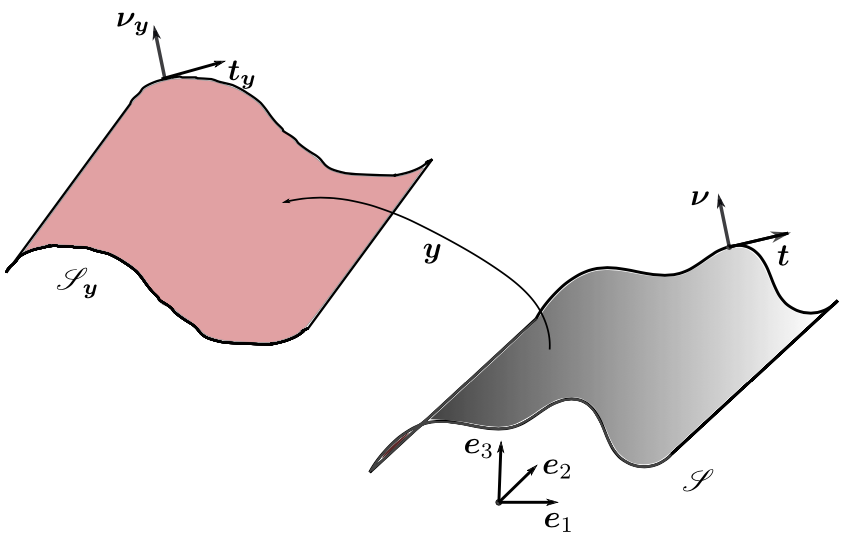}
	\caption{Deformation of cylindrical shells. Both $\surface$ and $\surface_{\y}$ are generated by the translation of a planar curve along $\e_2$; $(\tangent,\normal)$ and $(\tangent_{\y},\normal_{\y})$, both in the $(\e_1,\e_3)$ plane, are the unit tangent and normal vectors to the curves generating $\surface$ and $\surface_{\y}$, respectively.}
	\label{fig:cylinders}
\end{figure}

Here we consider the special class of \emph{cylindrical} shells, for which both $\surface$ and $\surface_{\y}$ are generated by the translation in space of  \emph{planar} curves along the direction orthogonal to their common plane: actually, it will suffice to know how $\y$ deforms the curve generating $\surface$ to obtain  $\surface_{\y}$. Figure~\ref{fig:cylinders}
shows both $\surface$ and $\surface_{\y}$ in a Cartesian frame $\framee$, where $\e_2$ is the direction of translation common to both surfaces; $\tangent$ is the unit tangent to the curve generating $\surface$, while $\tangent_{\y}$ is the unit tangent to the curve generating $\surface_{\y}$; $\normal$ and $\normal_{\y}$ are the corresponding unit normals.

According to \eqref{eq:deformation_gradient}, 
\begin{equation}
	\label{eq:cylinders_deformation_gradient}
	\nablas\y=\lambda\tangent_{\y}\otimes\tangent+\e_2\otimes\e_2,
\end{equation}
where $\lambda>0$ is the stretch of the generating curve. It readily follows from \eqref{eq:cylinders_deformation_gradient} that
\begin{equation}
	\label{eq:cylinders_U_R}
	\U=\lambda\tangent\otimes\tangent+\e_2\otimes\e_2\quad\text{and}\quad\R=\tangent_{\y}\otimes\tangent+\normal_{\y}\otimes\normal+\e_2\otimes\e_2.
\end{equation}
Moreover, letting $\kappa$ be the (signed) curvature of the curve generating $\surface$ and $\kappa_{\y}$ that of the curve generating $\surface_{\y}$, we have that
\begin{subequations}\label{cylinders_nablas}
	\begin{align}
		\nablas\tangent&=\kappa\normal\otimes\tangent,\quad\nablas\normal=-\kappa\tangent\otimes\tangent,\label{eq:cylinders_nabla_t_nabla_normal}\\
		\nablast\tangent_{\y}&=\kappa_{\y}\normal_{\y}\otimes\tangent_{\y},\quad\nablast\normal_{\y}=-\kappa_{\y}\tangent_{\y}\otimes\tangent_{\y},\label{eq:cylinders_nabla_ast_t_nabla_ast_normal}
	\end{align}
\end{subequations}
where, as already in \eqref{eq:connectors}, $\nablast$ denotes the surface gradient on $\surface_{\y}$. By the chain rule, equations \eqref{eq:cylinders_nabla_ast_t_nabla_ast_normal} imply that
\begin{equation}
	\label{eq:cylinders_nabla_t_y_nabla_normal_y}
	\nablas\tangent_{\y}=(\nablast\tangent_{\y})\nablas\y=\lambda\kappa_{\y}\normal_{\y}\otimes\tangent,\quad\nablas\normal_{\y}=(\nablast\normal_{\y})\nablas\y=-\lambda\kappa_{\y}\tangent_{\y}\otimes\tangent.
\end{equation}

Making use of \eqref{eq:cylinders_nabla_t_y_nabla_normal_y} in \eqref{eq:cylinders_U_R}, we obtain the representation for $\G$ in the form \eqref{eq:G_representation_alternative},
\begin{equation}
	\label{eq:cylinder_nablas_R}
	\G=(\kappa-\lambda\kappa_{\y})\W(\e_2)\otimes\tangent,
\end{equation}
where $\W(\e_2)$ is the skew-symmetric tensor associated with $\e_2$. It follows from \eqref{eq:cylinder_nablas_R} that 
\begin{subequations}\label{eq:cylinders_drilling_bending}
	\begin{align}
		\W(\normal)\circ\G&=(\kappa-\lambda\kappa_{\y})(\W(\normal)\cdot\W(\e_2))\tangent=2(\kappa-\lambda\kappa_{\y})(\normal\cdot\e_2)\tangent=\zero,\label{eq:cylinders_drilling}\\
		|\G|^2&=2(\lambda\kappa_{\y}-\kappa)^2.\label{eq:cylinders_bending}
	\end{align}
\end{subequations}
Thus, by \eqref{eq:w_d}, the pure measure of drilling $\wdr$ vanishes identically. This comes as no surprise, as drilling is inhibited in the class of cylindrical shells. For the same reason, the pure measure of bending $\wb$ need not vanish on the non-existent class of pure drilling deformations, and so it can be simply taken to be quadratic in $\G$,
\begin{equation}
	\label{eq:w_b_quadratic}
	\widetilde{w}_\mathrm{b}=|\G|^2,
\end{equation}
so that the energy density $W$ in \eqref{eq:W} can be replaced by 
\begin{equation}
	\label{eq:cylinder_W_quadartic}
	W=\frac12\mu_\mathrm{s}\ws+\frac12\widetilde{\mu}_\mathrm{b}\widetilde{w}_\mathrm{b}=\frac12\mu_\mathrm{s}(\lambda-1)^2+\widetilde{\mu}_\mathrm{b}(\lambda\kappa_{\y}-\kappa)^2,
\end{equation}
which is just the elastic energy posited by Antman~\cite{antman:general} for an elastic rod. It was already remarked in \cite{virga:pure} that $	\widetilde{w}_\mathrm{b}$ in \eqref{eq:w_b_quadratic} is a pure measure of bending. The interested reader is also referred to the abundant bibliography cited in \cite{hanna:assorted} for sources for an ampler discussion.





%

\end{document}